\documentclass[12pt, draftclsnofoot, onecolumn]{IEEEtran}
\usepackage{balance}
\usepackage{cite, graphicx,amssymb,color, pict2e, multirow, rotating}
\usepackage{graphicx,amssymb}
\usepackage[tbtags]{amsmath}
\usepackage{times,amsmath}
\usepackage{subfig}
\usepackage{flushend}
\usepackage{amsmath}
\usepackage{epsfig}
\usepackage{amssymb}
\usepackage{hyperref}
\usepackage{color}
\usepackage{psfrag}
\usepackage{subfig}
\usepackage[usenames,dvipsnames]{xcolor}
\usepackage{textcomp}
\usepackage{array}
\usepackage{tikz}
\usetikzlibrary{matrix,decorations.pathreplacing}
\usepackage{tabularx,ragged2e}
\newcolumntype{C}{>{\Centering\arraybackslash}X} 

\usepackage{algorithmic}
\usepackage{algorithm}

\newtheorem{definition}{Definition}

\newtheorem{lemma}{Lemma}

\newtheorem{theorem}{Theorem}

\title{The Study of Dynamic Caching via State Transition Field - the Case of Time-Varying Popularity}

\author{\IEEEauthorblockN{Jie~Gao, \IEEEmembership{Member, IEEE},   
		Lian~Zhao, \IEEEmembership{Senior Member, IEEE}, 
		and Xuemin~(Sherman)~Shen, \IEEEmembership{Fellow, IEEE} 
	}
	
		\thanks{
		J. Gao and X. Shen are with the Department of Electrical and Computer Engineering, University of Waterloo, Waterloo, ON, N2L 3G1, Canada (e-mail: \{jie.gao, sshen\}@uwaterloo.ca). 

		L. Zhao is with the Department of Electrical, Computer, and Biomedical Engineering, Ryerson University, Toronto, ON, M5B 2K3, Canada (e-mail: l5zhao@ryerson.ca).
		}
}
\vspace{-2mm}
\begin{document}

\maketitle

\vspace{-18mm}
\begin{abstract}
In the second part of this two-part paper, we extend the study of dynamic caching via state transition field (STF) to the case of time-varying content popularity. The objective of this part is to investigate the impact of time-varying content popularity on the STF and how such impact accumulates to affect the performance of a replacement scheme. Unlike the case in the first part, the STF is no longer static over time, and we introduce instantaneous STF to model it. Moreover, we demonstrate that many metrics, such as instantaneous state caching probability and average cache hit probability over an arbitrary sequence of requests, can be found using the instantaneous STF. As a steady state may not exist under time-varying content popularity, we characterize the performance of replacement schemes based on how the instantaneous STF of a replacement scheme after a content request impacts on its cache hit probability at the next request. From this characterization, insights regarding the relations between the pattern of change in the content popularity, the knowledge of content popularity exploited by the replacement schemes, and the effectiveness of these schemes under time-varying popularity are revealed. In the simulations, different patterns of time-varying popularity, including the shot noise model, are experimented. The effectiveness of example replacement schemes under time-varying popularity is demonstrated, and the numerical results support the observations from the analytic results. 
\end{abstract}


\section{Introduction}\label{s:intro} 

Driven by the upsurge in the number of user devices and their demand for multimedia services, the role of caching in improving the content delivery performance of wireless networks becomes prominent \cite{M_XWang2014} - \cite{C_JGao2018}. Accordingly, the modeling and analysis of caching have gained tremendous research attention~\cite{J_SMuller2017}-\cite{J_KLi2018}. While the independence reference model (IRM) is the de facto model for content requests, it has been argued that the IRM may not be sufficiently accurate in practice since temporal correlation of content requests can be too important to neglect \cite{M_GPaschos2016}. As a result, one particular topic, i.e., online caching with time-varying content popularity, has attracted great research interest lately  \cite{J_GSPaschos2018},~\cite{J_SAzimi2018}. 

The above-mentioned temporal correlation of content requests is sometimes referred to as `temporal locality', which suggests that a recently requested content is likely to be requested again in the near future. Temporal locality, however, has been shown to emerge from the temporal correlation of requests, the content popularity, or both~\cite{R_SJin1999}. Therefore, temporal locality exists even with IRM, and time-varying content popularity complicates the locality by introducing the temporal correlation. As a result, the study of online caching in the case of time-varying content popularity can be very challenging~\cite{J_GSPaschos2018}.  Existing research on caching with time-varying content popularity can be roughly categorized into two groups: the first group of works aims to analyze or model temporal locality, and the second group targets at proposing caching solutions to cope with it.      

Some early works on analyzing temporal locality focused on understanding its sources and developing metrics to measure it, e.g.,~\cite{C_MBusari2001}. In a recent work, Zhou~\textit{et~al.} investigated the change of popularity over time in the video-on-demand services~\cite{J_YZhou2015}. While the above studies tend to be experiment-based, mathematical models for characterizing temporal locality can be found in a few works. An inter-reference gap model was developed in~\cite{C_VPhalke1995}, which focused on describing temporal locality based on the gaps between successive requests. Traverso~\textit{et~al.} proposed a shot noise model~\cite{J_STraverso2013}, which represents the requests for a content with an inhomogeneous Poisson process, and later applied it on the analysis of video-on-demand traffic~\cite{J_STraverso2015}. Other approaches to integrate temporal locality into the analysis of caching also exist, most of which modeled the request for each content as a (semi-)Markov-modulated process or a renewal process~\cite{J_MGaretto2016},~\cite{C_MGaretto2015}.

By comparison, a larger number of works can be found in the second group, which proposes caching solutions to cope with temporal locality. Such solutions generally require the prediction of locality or the learning of content popularity. A cache replacement scheme based on predicting the interval between requests was proposed in \cite{J_AJaleel2010} and shown to be effective in increasing cache hits. Li~\textit{et~al.} developed a popularity-driven cache replacement scheme which learns the content popularity in an online fashion and makes replacement decisions based on the popularity forecast\cite{C_SLi2016}. Zhang~\textit{et~al.} proposed a model-free reinforcement learning algorithm for cache replacement based on a linear content popularity prediction model~\cite{C_NZhang2016}. The above works can be labeled as online caching based on learning/prediction since decisions for cache update are made after every content request. Another type of solutions is proactive caching based on prediction, which can handle time-varying content popularity assuming that cached contents are updated with a sufficiently high frequency. Sadeghi~\textit{et~al.} exploited reinforcement learning to track content popularity in an online fashion and developed a Q-learning based algorithm for content placement~\cite{J_ASadeghi2018}. Applegate~\textit{et~al.} formulated content placement as an optimization problem and, through estimating content popularity, proposed strategies to update cache contents to track time-varying content popularity~\cite{J_DApplegate2016}.  Bharath~\textit{et~al.} characterized the performance of caching with non-stationary content popularity from a learning-theoretic perspective and proposed a cache update policy based on the estimation of content popularity~\cite{J_BBharath2018}.    

Evidently, understanding the impact of time-varying content popularity on the performance of caching is important for the analysis and design of cache replacement schemes. However, analysis regarding the impact of time-varying popularity on the performance of replacement schemes is limited in the existing literature. The state transition field (STF) that we proposed in \cite{J_JGaoPartI2018} can be used for such analysis. However, with time-varying content popularity, the STF is no longer a static field but a dynamically varying field, and, consequently, a steady state may not exist. The objective of the second part of this two-part paper is to investigate the impact of time-varying content popularity on the STF and, as a result, the performance of replacement schemes. 

The contributions of the second part are the followings.  
         
First, we extend the concept of STF from the first part of this two-part paper \cite{J_JGaoPartI2018} and introduce instantaneous STF to characterize replacement schemes in the case of time-varying content popularity. It is shown that many metrics, such as instantaneous state caching probability (SCP) at an arbitrary instant and average cache hit probability over an arbitrary sequence of requests, can be found based on instantaneous STF. The results demonstrate the importance of instantaneous STF in modeling and analyzing replacement schemes with time-varying content popularity.

Second, as steady states may not exist, we characterize performance of a replacement scheme by analyzing the difference in instantaneous cache hit probability with and without applying that scheme after a content request. The result reveals insights regarding the relation between the change pattern in content popularity and the effectiveness of replacement schemes. We illustrate the results in the vector space of SCPs and relate them to the knowledge of content popularity exploited by replacement schemes.    

Third, we demonstrate instantaneous STF and average cache hit ratio under time-varying popularity with extensive simulations using example schemes. For instantaneous STF, we illustrate its relation with instantaneous content popularity and instantaneous cache hit probability. For  average cache hit ratio, we adopt different models of time-varying content popularity, including the shot noise model, and compare the performance of the example schemes. The results verify the observations from analysis and provide guidelines for designing replacement schemes under time-varying content popularity.

\section{System Model under Time-Varying Content Popularity}\label{s:sys}

For the sake of presentation clarity, we reintroduce some formulations from the first part of this two-part paper in Sections~\ref{s:sys}~and~\ref{s:ReplaceModel}. The basic system model follows from the basic model in the first part~\cite{J_JGaoPartI2018}. As the content popularity becomes time-varying, the symbols used here can be categorized into three groups based on their dependence on the time instant of content request or replacement: 
\begin{itemize}
	\item [G-1:] independent in both~\cite{J_JGaoPartI2018} and this paper;
	\item [G-2:] independent in~\cite{J_JGaoPartI2018} but dependent in this paper; 
	\item [G-3:] dependent in~\cite{J_JGaoPartI2018} and temporal locality introduces further dependence on the time instant in this paper;
\end{itemize}
A superscript $(\cdot)^{(n)}$ is added on symbols in groups G-2 and G-3 to denote the time instant related to the $n$th content request or replacement.

\subsection{Request-independent Symbols}

\emph{Cache State Vector/Matrix}: the cache state vector $\mathbf{s}_k$ for state $k$ and the cache state matrix $\mathbf{C}_\mathrm{s} = [\mathbf{s}_1, \dots, \mathbf{s}_{N_\mathrm{s}}]$, where $N_\mathrm{s}$ is the number of cache states.

\emph{Neighboring States}: the set of neighbors $\mathcal{H}_k$ and the set of content-$l$ neighbors $\mathcal{H}_{k, l}$ of state $k$, for any $k$ and any $l \notin \mathcal{C}_k$, where $\mathcal{C}_k$ is the set of cached contents in state $k$. 

The above symbols are in group G-1.

\subsection{Request-dependent Symbols}

\emph{Content Request Probabilities}: the probability of content $l$ being requested at request instant $n$, denoted by $\upsilon_l^{(n)}$, and the overall content popularity at the $n$th content request, denoted by $\boldsymbol{\upsilon}^{(n)}$. The content request probabilities are in symbol group G-2.

\emph{Instantaneous Cache Hit Probability}: the instantaneous cache hit probability at the $(n+1)$th request, denoted by $\gamma^{(n+1)}$ is given by:
\begin{align}
\gamma^{(n+1)} = \left(\boldsymbol{\upsilon}^{(n+1)} \right)^\mathrm{T} \boldsymbol{\lambda}^{(n)},
\end{align}
where $\cdot^\mathrm{T}$ represents transpose, and $\boldsymbol{\lambda}^{(n)}$ is the content caching probability (CCP) vector after the $n$th round of request and replacement. It can be seen that $\gamma^{(n+1)}$ is in symbol group G-3. Note that, in a practical network, there can be different metrics for content delivery, e.g., latency. However, the cache hit probability is an underlying factor which other metrics are dependent on. Consequently, improving the cache hit probability can improve the performance under other metrics. For example, if the cache hit probability at an edge server increases, then the need for retrieving contents from the cloud, and thus the average content delivery latency, reduces. Therefore, our study centers around the cache hit probability.

\emph{Station Transition Matrices}: The conditional state transition matrix and the state transition matrix are generally time dependent and thus denoted by $\boldsymbol{\Theta}_l^{(n)}$ and $\boldsymbol{\Theta}^{(n)}$, respectively, under time-varying content popularity. However, the situation is complicated by the possible choices of various replacement schemes and will be analyzed in details in Section~\ref{s:ReplaceModel}. 
 
It is worth noting that the relation between state and content caching probabilities from the first part of this two-part paper \cite{J_JGaoPartI2018}, i.e.,
\begin{align}\label{e:SCP2CCP} 
\boldsymbol{\lambda}^{(n)} = \mathbf{C}_\mathrm{s} \boldsymbol{\eta}^{(n)},  
\end{align}
still applies in the second part, where $\boldsymbol{\eta}^{(n)}$ is the SCP vector after the $n$th round of request and replacement. The above equation can be rewritten as:
\begin{align} \label{e:CCP2SCP} 
\boldsymbol{\eta}^{(n)} = \mathbf{C}_\mathrm{s}^\mathrm{T} \left(\mathbf{C}_\mathrm{s} \mathbf{C}_\mathrm{s}^\mathrm{T} \right)^{-1} \boldsymbol{\lambda}^{(n)} + \mathbf{n}_C^{(n)} ,  
\end{align}
where $\mathbf{n}_C^{(n)}$ can be any vector in the null space of $\mathbf{C}_\mathrm{s}$ that renders $\boldsymbol{\eta}^{(n)}$ a valid probability vector, i.e., $\boldsymbol{\eta}^{(n)} \succeq 0, \boldsymbol{\eta}^{(n)} \preceq 1$, and $\mathbf{1}^\mathrm{T} \boldsymbol{\eta}^{(n)} = 1$. Therefore, the value of $\mathbf{n}_C^{(n)}$ is dependent on the value of $\boldsymbol{\lambda}^{(n)}$. 

The content and state caching probabilities $\boldsymbol{\lambda}^{(n)}$ and $\boldsymbol{\eta}^{(n)}$ belong to group G-3 and will be analyzed in details in Section~\ref{s:CCPandSTF}.

\section{General Replacement Model and Specific Cases}\label{s:ReplaceModel}

In this section, the state transition probability matrix of the general replacement model is formulated, followed by the study of the four example replacement schemes introduced in the first part of this two-part paper, i.e., random replacement (RR), replace less popular (LP), replace the least popular (TLP), and least-recently-used (LRU) \cite{J_JGaoPartI2018}. Based on the state transition probability matrices, the instantaneous STF is defined at the end of this section.

\subsection{General Replacement Model}

Similar to the case in the first part, the state transition probability matrix in the general model can be written as:
\begin{align}\label{e:ThetaGenOverall}
\boldsymbol{\Theta}^{(n)} =  \sum\limits_{l \in \mathcal{C}} \upsilon_l^{(n)} \boldsymbol{\Theta}_l^{(n)}.
\end{align} 
where $\mathcal{C}$ is the set of all contents, and the conditional cache state transition probability matrix given that content $l \notin \mathcal{C}_k$ is requested, i.e., $\boldsymbol{\Theta}_l^{(n)}$, is given by:
\begin{align}\label{e:ThetaGen}
\boldsymbol{\Theta}_l^{(n)}(m, k) 
\!=  \! \left\{
\begin{array}{ll}
\! 1, \; & \text{if} \; k = m   \;\text{and} \; l \in \mathcal{C}_k, \\
\! 1 - \!\!\! \sum\limits_{m \in \mathcal{H}_{k,l}} \! \phi_{l, e(k, m), k}, \; & \text{if} \;  k = m\; \text{and} \; l \notin \mathcal{C}_k, \\
\! \phi_{l, e(k, m), k}, \; & \text{if} \;\; m \in \mathcal{H}_{k,l}, \\ 
\! 0, \; &\text{otherwise},
\end{array}
\right.
\end{align}
where $\phi_{l, q, k}$ denotes the probability of replacing content $q$ with content $l$ given that the cache is at state $k$ and content $l$ is requested. Unlike the case with time-invariant content popularity, the conditional cache state transition probability matrix $\boldsymbol{\Theta}_l^{(n)}$ can be implicitly request-dependent as a result of $\phi_{l, q, k}$ being request-dependent. Consider the situation when $e(k, m)= q$ and content $q$ is less popular at instant $n$ but more popular at instant $n^\prime$ compared to content $l$, i.e., $\upsilon_q^{(n)} < \upsilon_l^{(n)}$ and $\upsilon_q^{(n^\prime)} > \upsilon_l^{(n^\prime)}$. Consequently, $\phi_{l, q, k}$ can be different at instants $n$ and $n^\prime$ if LRU, LP, or TLP is used, and thus $\boldsymbol{\Theta}_l^{(n)}(m, k)$ can be different from $\boldsymbol{\Theta}_l^{(n^\prime)}(m, k)$. Using LRU as an example, the probability of content $q$ being the LRU content can be different at instants $n$ and $n^\prime$. Therefore, $\boldsymbol{\Theta}_l^{(n)}(m, k)$ is implicitly request-dependent although the request index $\cdot^{(n)}$ does not appear in the right-hand side of eq.~\eqref{e:ThetaGen}.

\subsection{RR}

It is straightforward to see that the conditional cache state transition probability matrix $\boldsymbol{\Theta}_l(m, k)$ in the case of RR is request-independent and remains the same as that in the first part. The overall state transition probability matrix $\boldsymbol{\Theta}_\mathrm{RR}$, however, becomes dependent on $(n)$ through $\boldsymbol{\upsilon}^{(n)}$: 
\begin{align}\label{e: ThetaOverallDef}
\boldsymbol{\Theta}^{(n)}_{\mathrm{RR}} (m, k)
=   \left\{
\begin{array}{ll}
1 - L \phi \sum\limits_{l \notin \mathcal{C}_k} \upsilon_l^{(n)} , \; & \text{if} \;\;  k = m, \\
\phi \upsilon_{e(m, k)}^{(n)}, \; &\text{if} \;\; m \in \mathcal{H}_k, \\ 
0, \; & \text{otherwise}, \\
\end{array}
\right. 
\end{align}
where $\phi \in (0, 1/L]$ represents the conditional replacement probability that any specified cached content is replaced given that the requested content is not in the cache. 

\subsection{LP}

In LP, an existing content may be replaced by the new content after the $n$th request if the new content is more likely to be requested at the $(n+1)$th request. The case of LP can be complicated as it involves the prediction of content popularity. Denote the prediction of content popularity at the $(n+1)$th request as $\tilde{\boldsymbol{\upsilon}}^{(n+1)}$. 
Sort the states in a non-decreasing order based on the sum of predicted request probability of the cached contents, i.e., 
\begin{align}
\sum_{q \in \mathcal{C}_m} \tilde{\upsilon}_q^{(n+1)} \geq \sum_{q \in \mathcal{C}_k} \tilde{\upsilon}_q^{(n+1)}, \quad \text{if} \; m \geq k. 
\end{align}
The state transition probability matrix of LP is then given by:
\begin{align}
& \boldsymbol{\Theta}^{(n)}_\mathrm{LP}(m, k) \nonumber \\
& = \left\{
\begin{array}{ll}
\!\! \sum\limits_{q \in \mathcal{C}_{k}} \upsilon_{q}^{(n)} \!\!+\!\! \sum\limits_{l \in \tilde{\mathcal{C}}_{\bar{k}^\downarrow}} \upsilon_{l}^{(n)} \!\!+\!\! \sum\limits_{ l \in \tilde{\mathcal{C}}_{\bar{k}^\uparrow}} \! \upsilon_{l}^{(n)} (1 \! - \! \alpha), \; &\text{if} \;\;  m = k, \\
\!\!  \alpha \upsilon_{e(m, k)}^{(n)} \phi_{e(m,k), e(k,m), k}^{(n)}, \; & \hspace{-18mm} \text{if} \;\; m > k \; \text{and} \; m \in \mathcal{H}_k, \\
\!\! 0, \; &\text{otherwise}, \\
\end{array}
\right.  
\end{align}  
in which $\alpha$ is the parameter for controlling the replacement probability,
\begin{align}
\phi_{l,q, k}^{(n)} &= \frac{\tilde{\upsilon}_l^{(n+1)} - \tilde{\upsilon}_q^{(n+1)}}{ \sum\limits_{ \{t| t \in \mathcal{C}_k, \tilde{\upsilon}_{t}^{(n+1)} < \tilde{\upsilon}_{l}^{(n+1)} \}} (\tilde{\upsilon}_l^{(n+1)} - \tilde{\upsilon}_t^{(n+1)}) },
\end{align}
and
\begin{subequations}
\begin{align}
\tilde{\mathcal{C}}_{\bar{k}^\downarrow} &= \left\{l \mid l \notin \mathcal{C}_{k}, \tilde{\upsilon}_l^{(n+1)} \leq \min_{t \in \mathcal{C}_k} \{ \tilde{\upsilon}_t^{(n+1)}\} \right\},  \\
\tilde{\mathcal{C}}_{\bar{k}^\uparrow} &= \left\{l \mid l \notin \mathcal{C}_{k}, \tilde{\upsilon}_l^{(n+1)} \geq \min_{t \in  \mathcal{C}_k} \{ \tilde{\upsilon}_t^{(n+1)}\} \right\}.
\end{align}
\end{subequations}

Note that the prediction $\tilde{\boldsymbol{\upsilon}}^{(n+1)}$ is not necessarily updated for each content request, and, as a result,  $\tilde{\boldsymbol{\upsilon}}^{(n+1)}$ can be a constant for a number of requests. The above state transition probability matrix applies regardless of what the predicted popularity stands for (i.e., the prediction can be for the next request or for a time period over multiple requests, etc.).

\subsection{TLP}

%
%

In TLP, an existing content is replaced after the $n$th request if it is both: i) the least likely to be requested among the cached content at the $(n+1)$th request; and ii) less likely to be requested at the $(n+1)$th request compared to the new content at the $n$th request. Sort the states in a non-decreasing order based on the sum of predicted request probability of the cached contents. The state transition probability matrix of TLP is given by:
\begin{align}\label{e: ThetaOverallDefLEA}
& \boldsymbol{\Theta}^{(n)}_\mathrm{TLP}(m, k) \nonumber \\
& = \! \left\{
\begin{array}{ll}
\!\! \sum\limits_{q \in \mathcal{C}_{k}} \!\! \upsilon_{q}^{(n)} \! + \!\!\sum\limits_{l \in \tilde{\mathcal{C}}_{\bar{k}^\downarrow}} \!\! \upsilon_{l}^{(n)} \!+ \!\! \sum\limits_{ l \in \tilde{\mathcal{C}}_{\bar{k}^\uparrow}} \!\! \upsilon_{l}^{(n)} (1 \!-  \! \phi_{l,q^\dagger(k), k}),  & \hspace{-2mm} \text{if} \;  m = k, \\
\!\! \upsilon_{e(m, k)} \phi_{e(m,k),q^\dagger(k), k}^{(n)}, \; & \hspace{-27mm} \text{if} \; m > k \; \text{and} \; k \in \mathcal{H}_{m, q^\dagger(k)}, \\
\!\! 0, \;& \hspace{-2mm}\text{otherwise}. \\
\end{array}
\right.  
\end{align}  
where $\phi_{l,q^\dagger(k), k}^{(n)}$ is the conditional probability of replacing $q^\dagger(k)$ with $l$ in state $k$, and
\begin{align}
q^\dagger(k) = \mathop{\mathrm{argmin}}\limits_{t \in \mathcal{C}_k} \{\tilde{\upsilon}_t^{(n+1)}\}. 
\end{align}
Note that $q^\dagger(k)$ changes over time although the superscript $\cdot^{(n)}$ is neglected here for simplicity of denotations. The value of $\phi_{e(m,k),q^\dagger(k), k}^{(n)}$, where $m > k$ and $k \in \mathcal{H}_{m, q^\dagger(k)}$, can be either 1 or $\tilde{\upsilon}_{e(m,k)}^{(n+1)} - \tilde{\upsilon}_{q^\dagger(k)}^{(n+1)}$, referred to TLP-A (always replace) and TLP-P (probabilistically replace), respectively. 

Similar to the case in the first part, $\boldsymbol{\Theta}^{(n)}_\mathrm{LP}$ and $\boldsymbol{\Theta}^{(n)}_\mathrm{TLP}$ are both lower-triangular matrices.

The relation among the content popularity, the prediction, and the SCP, all of which are time varying, can be very complicated. As our focus is on understanding the impact of replacement schemes on the  time-varying SCP instead of predicting content popularity, the prediction in the case of LP and TLP will be assumed to be accurate in this work. Same as in the first part, LP and TLP, unlike RR and LRU, are not practical replacement schemes but considered here just for analyzing the impact of content popularity information on the STF of replacement schemes.

\subsection{LRU}

To fit the LRU into the general cache state transition model, the conditional probability that a specific cached content is the LRU given the current cache state needs to be found. In order to find this conditional probability, the following result is obtained. 

\begin{lemma}\label{l:JointPLRU}	
	The joint probability that: i)~the current state is $k$; ii)~content $q^\star \in \mathcal{C}_k$ is the LRU content at the $n$th request; and iii)~the most recent request for $q^\star$ is the $(n-w)$th request,  denoted by $\rho^{(n)}(q^\star, w, k)$, can be found by:
	\begin{align}\label{e:lemmajointP} 
	\rho^{(n)}(q^\star, w, k)= 
	\sum\limits_{u = 1}^{U_w} \prod\limits_{i = 1}^{L- 1} \prod_{t \in \mathcal{T}(k, i, u, \bar{q}^\star)} \upsilon_{k(i, \bar{q}^\star)}^{(t)}.
	\end{align}
\end{lemma}
where $k(i, \bar{q}^\star), i \in \{1, \dots, L-1\}$ represents the $i$th cached content in state $k$ that is not content $q^\star$, $U_w$ represents the number of all possible ways for ordering and allocating $w-1$ requests to $L-1$ contents while guaranteeing at least one request for each content, and $\mathcal{T}(k, i, u, \bar{q}^\star)$ represents the set of requests allocated to content $k(i, \bar{q}^\star)$ in the $u$th out of the $U_w$ allocations.

\textit{Proof}: See Section~\ref{p:JointPLRU} in Appendix.

Given the joint probability in Lemma~\ref{l:JointPLRU}, the conditional probability that content $q^\star\in \mathcal{C}_k$ is the LRU content given that the current state is $k$ can be found as follows~\footnote{Here it is assumed that a sufficient number of requests have occurred, i.e., $n\rightarrow \infty$.}:
\begin{align}
\rho^{(n)}(q^\star| k) = \frac{\sum\limits_{w = L}^{\infty} \rho^{(n)}(q^\star, w, k)}{\sum\limits_{w = L}^{\infty} \sum\limits_{q \in \mathcal{C}_k} \rho^{(n)}(q, w, k) }.
\end{align}
Note that the above probability is the general case for the probability $\rho^\text{LRU}_{e(k,m)|k}$ from the first part of this two-part paper.

Using the above conditional probability, the conditional state transition probability matrix $\boldsymbol{\Theta}_l$ can be given by:
\begin{align}\label{e: ThetaDefLRU}
\boldsymbol{\Theta}_{l, \mathrm{LRU}}^{(n)}(m, k) 
=   \left\{
\begin{array}{ll}
1, \; & \text{if} \;\;  l \in \mathcal{C}_k  \; \text{and} \;  k = m,  \\
\rho^{(n)}(e(k,m)| k),\; & \text{if} \;\; m = \mathcal{H}_{k,l}, \\ 
0, \; & \text{otherwise}. \\
\end{array}
\right.
\end{align}
The overall state transition probability matrix $\boldsymbol{\Theta}_\mathrm{LRU}$ is given by:
\begin{align}\label{e: ThetaOverallDefLRU}
\boldsymbol{\Theta}_\mathrm{LRU}^{(n)} (m, k)
=   \left\{
\begin{array}{ll}
\sum\limits_{l \in \mathcal{C}_k} \upsilon_l^{(n)}, \; & \text{if} \;\;  k = m, \\
\upsilon_{e(m, k)}^{(n)} \rho^{(n)}(e(k,m)| k), \; & \text{if} \;\; m \in \mathcal{H}_k, \\ 
0, \; & \text{otherwise}. \\
\end{array}
\right. 
\end{align}

\section{Instantaneous CCP and STF}\label{s:CCPandSTF}

Based on the state transition probability matrix, this section analyzes the transition of the instantaneous CCP and formulates the instantaneous STF.

\subsection{Instantaneous CCP}\label{s:CCP}

Based on the relation between the content and the state caching probabilities in eq.~\eqref{e:SCP2CCP}, the resulting CCP vector after the $n$th request and replacement is given by:
\begin{align}
\boldsymbol{\lambda}^{(n)} =  \mathbf{C}_\mathrm{s} \boldsymbol{\eta}^{(n)} = \mathbf{C}_\mathrm{s} \sum\limits_{l \in \mathcal{C}} \upsilon_l^{(n)} \boldsymbol{\Theta}_l^{(n)} \boldsymbol{\eta}^{(n-1)}. 
\end{align}
Using eq.~\eqref{e:CCP2SCP}, it follows that:
\begin{align}\label{e:lambdaOneStep}
\boldsymbol{\lambda}^{(n)} 
=& \left(\mathbf{C}_\mathrm{s} \sum\limits_{l \in \mathcal{C}} \upsilon_l^{(n)} \boldsymbol{\Theta}_l^{(n)} \mathbf{C}_\mathrm{s}^\mathrm{T} \left(\mathbf{C}_\mathrm{s} \mathbf{C}_\mathrm{s}^\mathrm{T} \right)^{-1}\right) \boldsymbol{\lambda}^{(n-1)} \nonumber \\ & + \mathbf{C}_\mathrm{s} \sum\limits_{l \in \mathcal{C}} \upsilon_l^{(n)} \boldsymbol{\Theta}_l^{(n)} \mathbf{n}_C^{(n-1)}.
\end{align}

It can be seen that the mapping from $\boldsymbol{\lambda}^{(n-1)}$ to $\boldsymbol{\lambda}^{(n)}$ is complicated. Specifically, unlike the mapping between two consecutive SCP vectors, which can be simply written as $\boldsymbol{\eta}^{(n)} = \mathbf{\Theta}^{(n)} \boldsymbol{\eta}^{(n-1)}$, the mapping between consecutive CCP vectors cannot be written in a linear form due to the second item in eq.~\eqref{e:lambdaOneStep}, i.e., $\mathbf{C}_\mathrm{s} \sum_{l \in \mathcal{C}} \upsilon_l^{(n)} \boldsymbol{\Theta}_l^{(n)} \mathbf{n}_C^{(n-1)}$. Moreover, despite that eq.~\eqref{e:lambdaOneStep} seems to have an affine form, the mapping from $\boldsymbol{\lambda}^{(n-1)}$ to $\boldsymbol{\lambda}^{(n)}$ is not affine either. This is implicitly conveyed through the variable $\mathbf{n}_C^{(n-1)}$ since the value of $\mathbf{n}_C^{(n-1)}$ depends on $\boldsymbol{\lambda}^{(n-1)}$ and the dependence is nonlinear as explained after eq.~\eqref{e:CCP2SCP} in Section~\ref{s:sys}.

\subsection{Instantaneous STF - The General Case}


Under time-varying content popularity, the state transition probability matrix is $\boldsymbol{\Theta}^{(n)}$ when the SCP is $\boldsymbol{\eta}^{(n-1)}$. Therefore, the STF at the instant of the $n$th request and the point $\boldsymbol{\eta}^{(n-1)}$ is given by:
\begin{align}\label{e:STfield}
\boldsymbol{u}^{(n)}(\boldsymbol{\eta}^{(n-1)}) = \boldsymbol{\Theta}^{(n)} \boldsymbol{\eta}^{(n-1)} - \boldsymbol{\eta}^{(n-1)}.
\end{align}
The superscript $(n)$ in $\boldsymbol{u}^{(n)}(\cdot)$ reflects the fact that the STF is no longer static but time-varying as a result of the time-varying content popularity. The direction and strength of the instantaneous STF depend on both $\boldsymbol{\eta}$, i.e., the location in the state transition domain, and $n$, i.e., the request instant. The value of the instantaneous STF $\boldsymbol{u}^{(n)}(\boldsymbol{\eta}^{(n-1)}) $ represents the change in the SCP after the $n$th round of request and replacement. The effect of a replacement scheme on the dynamic SCP over a sequence of requests can be decomposed into the summation over the instantaneous STFs:
\begin{align}\label{e:etaSequenceDecomp}
\boldsymbol{\eta}^{(n + N -1)} - \boldsymbol{\eta}^{(n -1)}  
&= \sum\limits_{t = 0}^{N-1} \left(\boldsymbol{\eta}^{(n + t)} - \boldsymbol{\eta}^{(n + t - 1)} \right) \nonumber\\
&=  \sum\limits_{t = 0}^{N-1} \boldsymbol{u}^{(n + t)}(\boldsymbol{\eta}^{(n+t-1)}),
\end{align} 
for any $n \geq 1$ and $N \geq 1$. 

Similarly, other metrics can also be studied through instantaneous STFs, e.g., the average cache hit probability.

\begin{lemma}\label{l:gammaSequence}
Using instantaneous STFs from the first till the $n$th request, the average cache hit probability over the $n$ requests can be given by:	
\begin{align}\label{e:gammaSequenceLemma}
\gamma_\textrm{avg} =  \frac{1}{n} \sum\limits_{t = 2}^{n} \left(\boldsymbol{\upsilon}^{(t)}\right)^\mathrm{T} \mathbf{C}_\mathrm{s} \bigg(\sum\limits_{t^\prime = 0}^{t-2} \boldsymbol{u}^{(t^\prime +1)} \bigg) +  \boldsymbol{\upsilon}_\mathrm{avg}^\mathrm{T} \mathbf{C}_\mathrm{s} \boldsymbol{\eta}^{(0)},
\end{align}
in which $\boldsymbol{u}^{(t^\prime +1)}$ is the abbreviation for $\boldsymbol{u}^{(t^\prime +1)}(\boldsymbol{\eta}^{(t^\prime)})$, and
\begin{align}
\boldsymbol{\upsilon}_\mathrm{avg} = \frac{1}{n}\sum\limits_{t = 1}^{n} \boldsymbol{\upsilon}^{(t)}
\end{align}
is the average content popularity over the $n$ requests. 
\end{lemma}

\text{Proof}: See Section~\ref{p:gammaSequence} in Appendix.

Lemma~\ref{l:gammaSequence} shows that the average cache hit probability over an arbitrary number of requests, starting from any initial SCP $\boldsymbol{\eta^{(0)}}$, can be obtained from instantaneous STFs, instantaneous content request probabilities, and the initial point $\boldsymbol{\eta^{(0)}}$.  The inner summation over $t^\prime$ in eq.~\eqref{e:gammaSequenceLemma} represents the effect of historical requests and replacements on the instantaneous cache hit probability at the $t$th request. The decomposition in eq.~\eqref{e:etaSequenceDecomp} and the result in eq.~\eqref{e:gammaSequenceLemma} demonstrate the importance in analyzing the instantaneous STF under different replacement schemes. If the instantaneous content request probabilities $\boldsymbol{\upsilon}^{(t)}, t \in \{1, \dots, n\}$ can be obtained, the instantaneous STF of a replacement scheme at any point in the state transition region can be calculated using eqs.~\eqref{e:ThetaGenOverall}, \eqref{e:ThetaGen}, and \eqref{e:STfield}. For evaluating and comparing different cache replacement schemes, we can substitute the specific STF of the replacement schemes for $\boldsymbol{u}^{(1)}, \dots, \boldsymbol{u}^{(t - 1)}$ in eq.~\eqref{e:gammaSequenceLemma}.

The instantaneous STF can be decomposed. Define the $l$th component of $\mathbf{u}^{(n)}(\boldsymbol{\eta}^{(n-1)})$ as:
\begin{align}\label{e:u_l}
\mathbf{u}_l^{(n)} =  \boldsymbol{\Theta}_l^{(n)} \boldsymbol{\eta}^{(n-1)} - \boldsymbol{\eta}^{(n-1)}. 
\end{align}
It can be seen that:
\begin{align}\label{e:u_decomp}
\mathbf{u}^{(n)}(\boldsymbol{\eta}^{(n-1)}) =& \boldsymbol{\Theta}^{(n)} \boldsymbol{\eta}^{(n-1)} - \boldsymbol{\eta}^{(n-1)} \nonumber\\
=& \sum\limits_{l \in \mathcal{C}} \upsilon_l^{(n)} \left( \boldsymbol{\Theta}_l\boldsymbol{\eta}^{(n-1)}  - \boldsymbol{\eta}^{(n-1)} \right) \nonumber\\
=& \sum\limits_{l \in \mathcal{C}} \upsilon_l^{(n)}  \mathbf{u}_l^{(n)}.
\end{align}

\subsection{The Case of RR, LP, TLP, and LRU}

When a specific replacement scheme is considered, $\mathbf{u}_l^{(n)}$ can be found based on its conditional state transition probability matrix $\boldsymbol{\Theta}^{(n)}_l$ using \eqref{e:u_l}. 

For the case of RR, the $m$th element of $\mathbf{u}_{l}^{(n)}$ is given by:
\begin{align}\label{e:ul_RR}
u_{m, l, \mathrm{RR}} =  \left\{
\begin{array}{ll}
\phi \sum\limits_{\{k| m \in \mathcal{H}_{k,l}\}} \eta_k, \; & \text{if}\;  l \in \mathcal{C}_m, \\ 
-L \phi \eta_m, \; & \text{otherwise}. 
\end{array}
\right.
\end{align}

The $m$th element of $\mathbf{u}_l^{(n)}$ for LP is given by:
\begin{align}
&u_{m, l, \mathrm{LP}}^{(n)} \nonumber \\  &=  \left\{
\begin{array}{ll}
\sum\limits_{k \in \mathcal{G}_{m, l}^{(n)}} \eta_k^{(n-1)} \phi_{l, e(k, m), k}^{(n)}, &\text{if}\; l \in \mathcal{C}_m, \\
- \eta_m^{(n-1)}, \; &\hspace{-2cm} \text{if}\; l \notin \mathcal{C}_m\, \text{and} \, \min\limits_{q \in \mathcal{C}_m} \{\tilde{\upsilon}_{q}^{(n+1)}\} < \tilde{\upsilon}_l^{(n+1)}, \\ 
0, \; &\hspace{-2cm} \text{otherwise}, 
\end{array}
\right.
\end{align}
where
\begin{align}
\mathcal{G}_{m, l}^{(n)} = \{ k | m \in \mathcal{H}_{k,l}, \tilde{\upsilon}_{e(k, m)}^{(n+1)} < \tilde{\upsilon}_l^{(n+1)} \},
\end{align}
representing the set of states which include state $m$ in their content-$l$ neighbors and cache a less popular content compared to state $m$ according to the predicted popularity for the $(n+1)$th request. 

Similarly, the $m$th element of $\mathbf{u}_l^{(n)}$ for TLP is given by:
\begin{align}
&u_{m, l, \mathrm{TLP}}^{(n)}  \nonumber \\  &=  \left\{
\begin{array}{ll}
\!\! \sum\limits_{k \in \hat{\mathcal{G}}_{m, l}^{(n)}}  \phi_{e(m,k),q^\dagger(k), k}^{(n)} \eta_k^{(n-1)}, &\text{if}\; l \in \mathcal{C}_m, \\
\!\! - \phi_{e(m,k),q^\dagger(k), k}^{(n)} \eta_m^{(n-1)}, \\ & \hspace{-2cm} \text{if}\; l \notin \mathcal{C}_m\, \text{and} \, \min\limits_{q \in \mathcal{C}_m} \{\tilde{\upsilon}_{q}^{(n+1)}\} < \tilde{\upsilon}_l^{(n+1)}, \\ 
\!\! 0, &  \text{otherwise}, 
\end{array}
\right.
\end{align}
where 
\begin{align}
\hat{\mathcal{G}}_{m, l}^{(n)} = \{ k | m \in \mathcal{H}_{k,l}, \tilde{\upsilon}_{e(k, m)}^{(n+1)} =  \min\limits_{q \in \mathcal{C}_k} \{\tilde{\upsilon}_{q}^{(n+1)}\} < \tilde{\upsilon}_l^{(n+1)} \},
\end{align}
representing the set of states which include state $m$ in their content-$l$ neighbors and cache a content less popular than any content cached by state $m$ according to the predicted popularity for the $(n+1)$th request.

For the case of LRU, the $m$th element of $\mathbf{u}_l^{(n)}$ is given by:
\begin{align}\label{e:ul_LRU}
&u_{m, l, \mathrm{LRU}}^{(n)}  \nonumber \\  &=  \left\{
\begin{array}{ll}
\sum\limits_{k \in \mathcal{G}_{m, l}} \rho^{(n)}(e(k,m) |k)  \eta_k^{(n-1)}, \, & \text{if}\; l \in \mathcal{C}_m \\
- \eta_m^{(n-1)}, \; &\text{otherwise}\\ 
\end{array}
\right.
\end{align}
where
\begin{align}
\mathcal{G}_{m, l} = \{ k | m \in \mathcal{H}_{k,l} \}.
\end{align}

In the next section, we study the instantaneous STF of the considered replacement schemes and its impact on their instantaneous cache hit probability.

\section{Impact of STF on Instantaneous Cache Hit Probability}

When the content popularity varies over time, a replacement scheme may not lead to any steady state. As a result, the analysis of steady states and rate of convergence does not apply. Instead, the impact of a replacement scheme on the instantaneous cache hit probability at the next request is investigated. 

\subsection{The General Case} 

A replacement after the $n$th request affects the cache hit probability at the $(n+1)$th request. Consider the time instant right after the $n$th request and replacement so that $\boldsymbol{u}^{(n)}(\cdot)$ is the current STF and the $(n+1)$th request is the next request in future. The effect of a replacement scheme can be conveyed through the difference between the cache hit probability at the $(n+1)$th request with and without a replacement (based on the chosen scheme) after the $n$th request. This difference is given by:
\begin{align}\label{e:difHitRate_n+1}
 d^{(n+1)}_\gamma = & \left(\boldsymbol{\upsilon}^{(n+1)} \right)^\mathrm{T} \mathbf{C}_\mathrm{s}\left( \boldsymbol{\eta}^{(n)} -  \boldsymbol{\eta}^{(n-1)}\right) \nonumber \\
= & \left(\boldsymbol{\upsilon}^{(n+1)} \right)^\mathrm{T} \mathbf{C}_\mathrm{s}\boldsymbol{u}^{(n)}(\boldsymbol{\eta}^{(n-1)}).
\end{align}
The above result shows that, the cache hit ratio at the $(n+1)$th request depends on the content popularity at the $(n+1)$th request, i.e., $\boldsymbol{\upsilon}^{(n+1)}$, the STF at the $n$th request, i.e., $\boldsymbol{u}^{(n)}(\cdot)$, and the SCP at the $(n-1)$th request, i.e., $\boldsymbol{\eta}^{(n-1)}$. Among these three factors, $\boldsymbol{\eta}^{(n-1)}$ reflects the accumulative effect of the previous $n-1$ rounds of request and replacement, $\boldsymbol{u}^{(n)}(\cdot)$ represents the current STF, and $\boldsymbol{\upsilon}^{(n+1)}$ represents the content popularity at the next request in future. The result in eq.~\eqref{e:difHitRate_n+1} shows the complication due to time-varying content popularity: $\boldsymbol{\upsilon}^{(n+1)}$ and $\boldsymbol{u}^{(n)}(\cdot)$ in eq.~\eqref{e:difHitRate_n+1} would reduce to $\boldsymbol{\upsilon}$ and $\boldsymbol{u}(\cdot)$, respectively, if the content popularity becomes time-invariant.

\begin{figure}[tt]
	\centering {\includegraphics[width=0.45\textwidth]{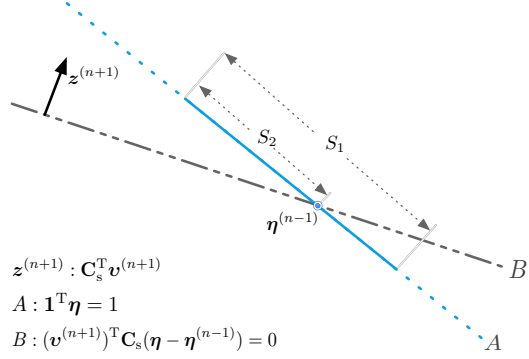}} 
	\vspace{-2mm}
	\caption{Illustration of the relation between instantaneous cache hit probability, $\boldsymbol{\eta}^{(n)}$, and $\boldsymbol{\upsilon}^{(n+1)}$. Area $S_1$ is the area that $\boldsymbol{\eta}^{(n+1)}$ may fall in, i.e., the intersection of hyperplane $A$ and the subspace $\boldsymbol{\eta}^{(n+1)} \succeq 0$. If $\boldsymbol{\eta}^{(n+1)}$ falls in area $S_2$, then $(\boldsymbol{z}^{(n+1)})^{T}\boldsymbol{\eta}^{(n+1)} \geq (\boldsymbol{z}^{(n+1)})^{T}\boldsymbol{\eta}^{(n)}$.}\label{f:OneStepImrpovement}
\end{figure}

Some general observations can be made:
\begin{enumerate}
	\item Define $\boldsymbol{z}^{(n+1)} = \mathbf{C}_\mathrm{s}^\mathrm{T}\boldsymbol{\upsilon}^{(n+1)}$. Then $\boldsymbol{z}^{(n+1)}$ is the state cache hit probability vector at the $(n+1)$th request. Depending on $\boldsymbol{\eta}^{(n-1)}$, $\boldsymbol{\upsilon}^{(n)}$, and $\boldsymbol{\Theta}^{(n)}$, $\boldsymbol{\eta}^{(n+1)}$ may fall at any point in the areas $S_1$ in Fig.~\ref{f:OneStepImrpovement}. The replacement after the $n$th request improves the instantaneous cache hit probability at the $(n+1)$th request if the replacement drives the SCP into the area $S_2$ shown in Fig.~\ref{f:OneStepImrpovement}.
	\item  $d^{(n+1)}_\gamma$ is small, regardless of $\boldsymbol{\upsilon}^{(n+1)}$, when $\boldsymbol{\eta}^{(n-1)}$ is close to the steady state corresponding to $\boldsymbol{\upsilon}^{(n)}$ (i.e., the steady state if the content popularity is constant and remains equal to $\boldsymbol{\upsilon}^{(n)}$). 
	\item In the trivial case when $\boldsymbol{\upsilon}^{(n+1)}$ approaches $1/N_\mathrm{c} \cdot \mathbf{1}$, where $N_\mathrm{c}$ is the number of contents,  
	 the hyperplane $(\boldsymbol{\upsilon}^{(n+1)})^T\mathbf{C}_\mathrm{s} (\boldsymbol{\eta} -  \boldsymbol{\eta}^{(n)}) = 0$ coincides with the hyperplane $\mathbf{1}^\mathrm{T} \boldsymbol{\eta} = 1$. In such case, $d^{(n+1)}_\gamma$ becomes zero for any replacement scheme. 
\end{enumerate}

\hspace{-12mm} 

The effect of a replacement scheme on $d^{(n+1)}_\gamma$ can be conveyed through the set of content-specific instantaneous STF $\{\mathbf{u}^{(n)}_l\}$ using eq.~\eqref{e:u_decomp}.
\begin{theorem}\label{t:dGammaTheo}
	The $d^{(n+1)}_\gamma$ in eq.\eqref{e:difHitRate_n+1} can be equivalently rewritten as:
	\begin{align}
	d^{(n+1)}_\gamma & =  \sum\limits_{l \in \mathcal{C}} (\upsilon_l^{(n)} - \bar{\upsilon}_l) c^{(n+1)}_l, \label{e:difHitRate_n+1_2}  
	\end{align}
	where 
	\begin{align}
	c_l^{(n+1)} &= \left(\boldsymbol{\upsilon}^{(n+1)} \right)^\mathrm{T} \mathbf{C}_\mathrm{s} \mathbf{u}_l^{(n)}, \label{e:alpha}
	\end{align}
	and $\{\bar{\upsilon}_l\}_{l \in \mathcal{C}}$ represents the content popularity under which $\boldsymbol{\eta}^{(n-1)}$ would be the steady state. 
\end{theorem}

\textit{Proof}: See Section~\ref{p:dGammaTheo} in Appendix.

Based on eq.~\eqref{e:difHitRate_n+1_2} and eq.~\eqref{e:alpha}, the factors that determine $d^{(n+1)}_\gamma$ are: $\{\upsilon_l^{(n+1)}\}$, $\{\upsilon_l^{(n)}\}$, $\{\bar{\upsilon}_l\}$, and $\mathbf{u}_l^{(n)}$. The factor $\{\bar{\upsilon}_l\}$ depends on the historical content requests till the $(n-1)$th request, $\mathbf{u}_l^{(n)}$ depends on $\boldsymbol{\eta}^{(n)}$, and both $\{\bar{\upsilon}_l\}$ and $\mathbf{u}_l^{(n)}$ depend on the replacement scheme. The term $\upsilon_l^{(n)} - \bar{\upsilon}_l$ reflects the deviation in the request probability for content $l$ from its `steady' request probability, which manifests the influence of historical requests. The term $c_l^{(n+1)}$ represents the change in the cache hit probability at the $(n+1)$th request, using the corresponding replacement scheme, when the current SCP is $\boldsymbol{\eta}^{(n-1)}$ and content $l$ is requested at the $n$th request.

Using Theorem~\ref{t:dGammaTheo}, a more detailed investigation could be conducted for a specific content popularity model (i.e., shot noise model \cite{J_STraverso2013}). Nevertheless, the study on specific content popularity models is not the focus of this work. Section~\ref{s:Simu}, however, will cover numerical results on the performance of replacement schemes under specific content popularity models.

\subsection{Upper and Lower Bounds of $d^{(n+1)}_\gamma$}

The term $\mathbf{C}_\mathrm{s}\boldsymbol{u}^{(n)}(\boldsymbol{\eta}^{(n-1)})$ in $d^{(n+1)}_\gamma$ represents the change in the content caching probabilities after the $n$th request under the chosen replacement scheme. Sort the contents based on their popularity at the instant of the $n$th request so that $ \upsilon_1^{(n)} \geq \upsilon_2^{(n)} \geq \dots \geq  \upsilon_{N_\mathrm{c}}^{(n)}$. The upper-bound and lower-bound of $d^{(n+1)}_\gamma$ can be found using the following result.

\begin{theorem}\label{t:dBound}
	The upper-bound and lower-bound of $d^{(n+1)}_\gamma$, denoted as $\hat{d}^{(n+1)}_\gamma$ and $\check{d}^{(n+1)}_\gamma$, respectively, for RR, LP, TLP, and LRU are given by~\footnote{For the lower-bound of $d^{(n+1)}_\gamma$ in the case of TLP-P, it is assumed that the $L$ least popular contents at the $n$th request remain least popular at the $(n+1)$th request.}: 
	\begin{align}
	\hat{d}^{(n+1)}_\gamma =  \left\{
	\begin{array}{ll}
	L \phi  \max\limits_l \{ \upsilon_l^{(n)} \},  &\text{RR}\\
	\alpha  \max\limits_l \{ \upsilon_l^{(n)} \},  &\text{LP}\\ 
	 \max\limits_l \{ \upsilon_l^{(n)} \},     &\text{TLP-A\, or\, LRU}\\
	\max\limits_l \{ \upsilon_l^{(n)} \}\max\limits_l \{\tilde{\upsilon}_l^{(n+1)}\}, \qquad &\text{TLP-P}
	\end{array}
	\right.
	\end{align}
	and
	\begin{align}
	\check{d}^{(n+1)}_\gamma =  \left\{
	\begin{array}{ll}
	-\phi,  &\text{RR}\\
	-\alpha,  &\text{LP}\\ 
	-1,   &\text{TLP-A \, or\, LRU} \\
	-\sum\limits_{l=1}^{N_\mathrm{c}} \upsilon_l^{(n)}\tilde{\upsilon}_l^{(n+1)},  &\text{TLP-P}.
	\end{array}
	\right.
	\end{align}
\end{theorem}

\textit{Proof}: See Section~\ref{p:dBound} in Appendix.

\begin{figure*}[t]
	\centering \subfloat[$\mathbf{u}^{(n)}$ and $\boldsymbol{z}^{(n+1)}$ : cases~1~and~2.]
	{\includegraphics[angle=0,width=0.39\textwidth]{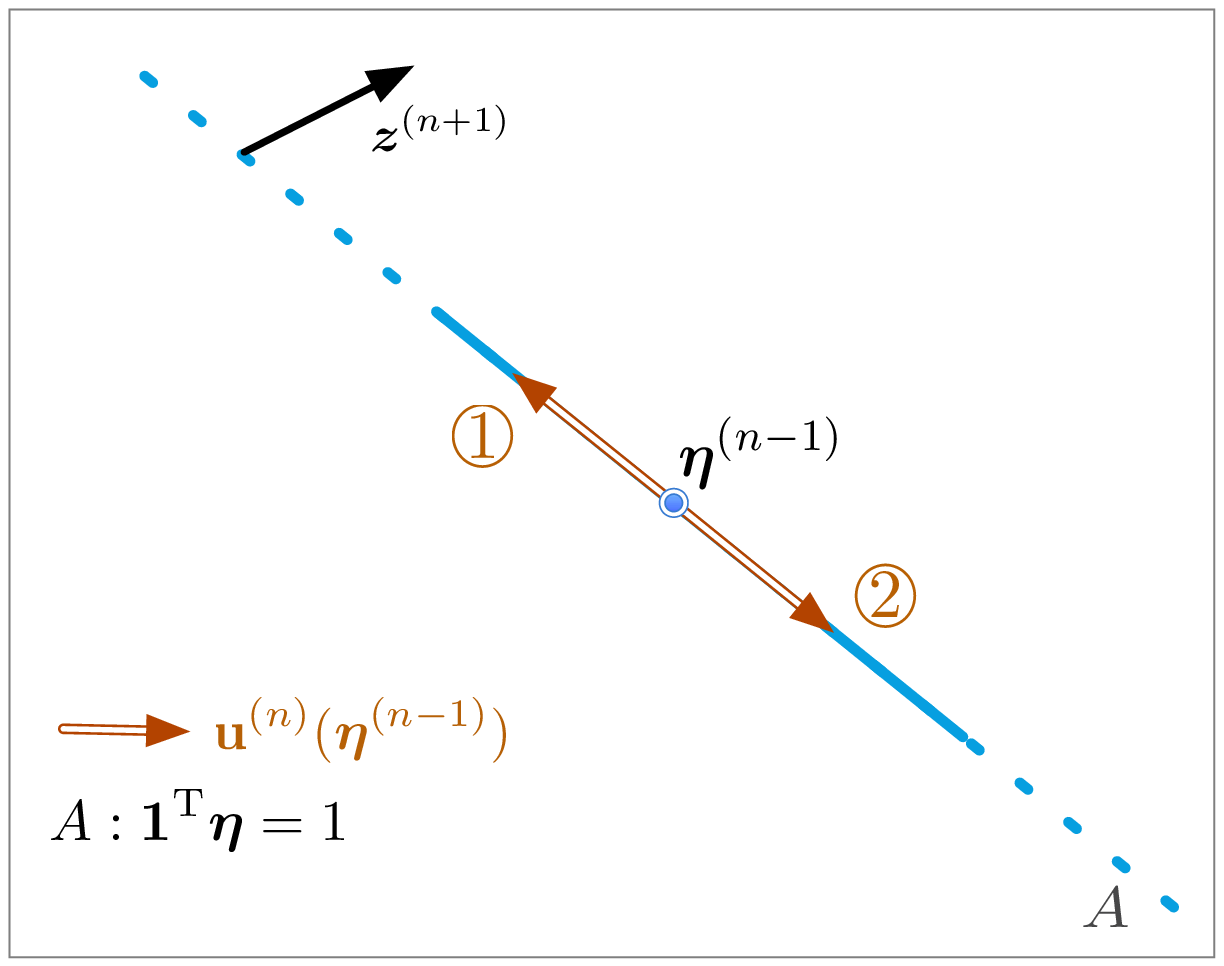} 
		\label{f:ulcase12}}
	\hspace{-1mm} 
	\subfloat[$\mathbf{u}^{(n)}$ and $\boldsymbol{z}^{(n+1)}$: cases~3~and~4.]
	{\includegraphics[angle=0,width=0.39\textwidth]{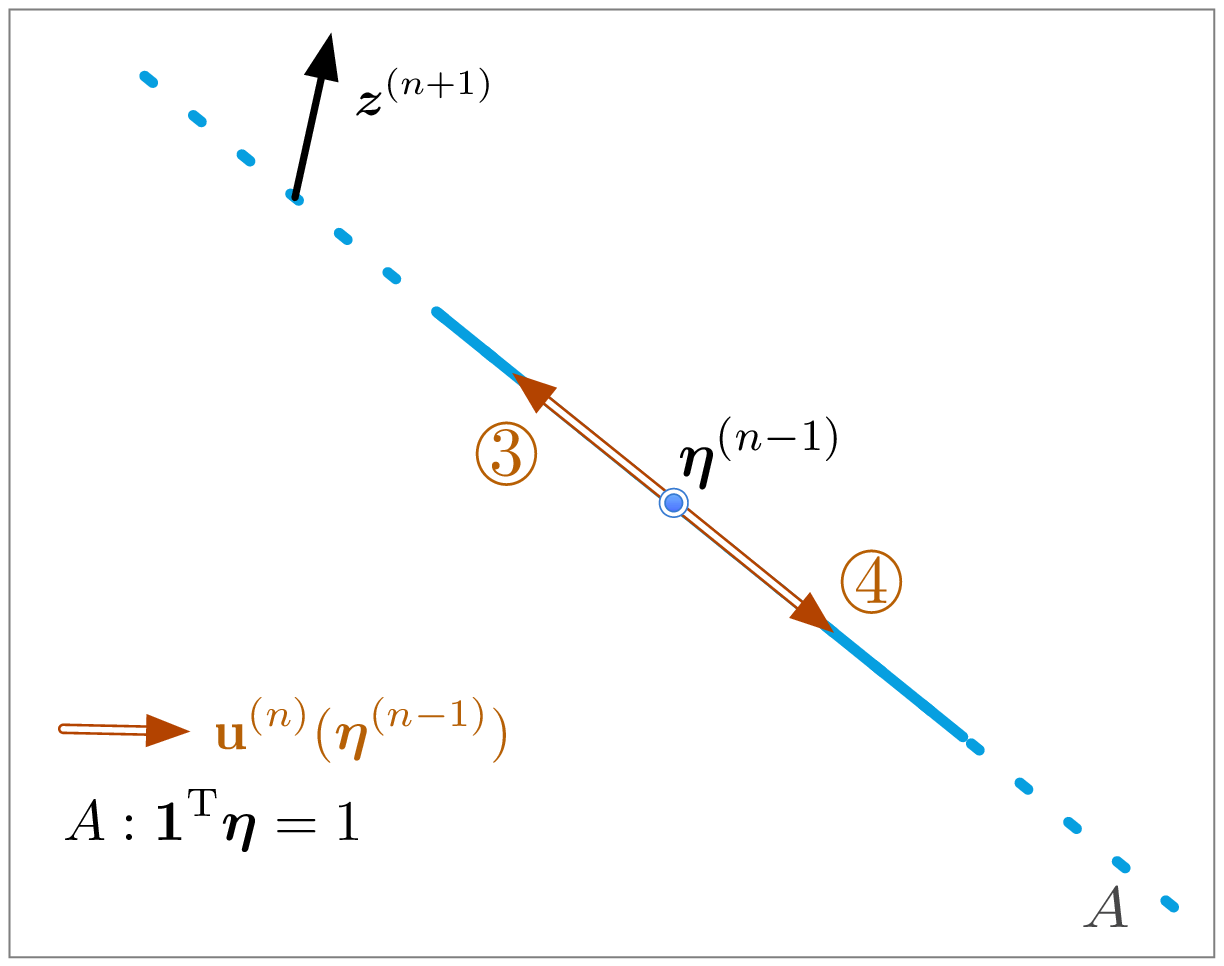} 
		\label{f:ulcase34}}
	\hspace{1mm} 
	\subfloat[$\boldsymbol{z}^{(n)}$ changes along a straight path.]
	{\includegraphics[angle=0,width=0.39\textwidth]{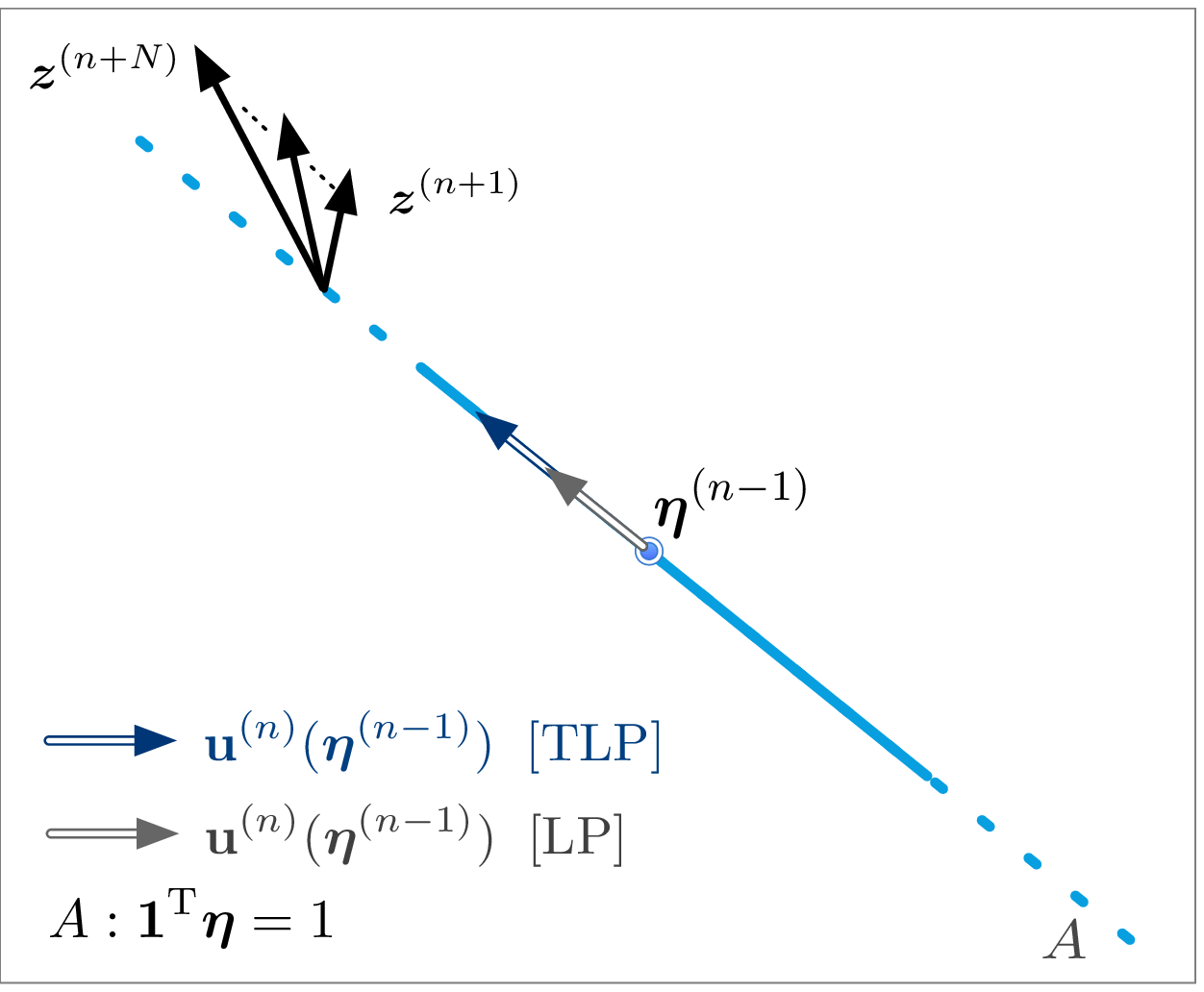} 
		\label{f:PopularityStraight}}
	\hspace{0mm} 
	\subfloat[$\boldsymbol{z}^{(n)}$ changes randomly.]
	{\includegraphics[angle=0,width=0.39\textwidth]{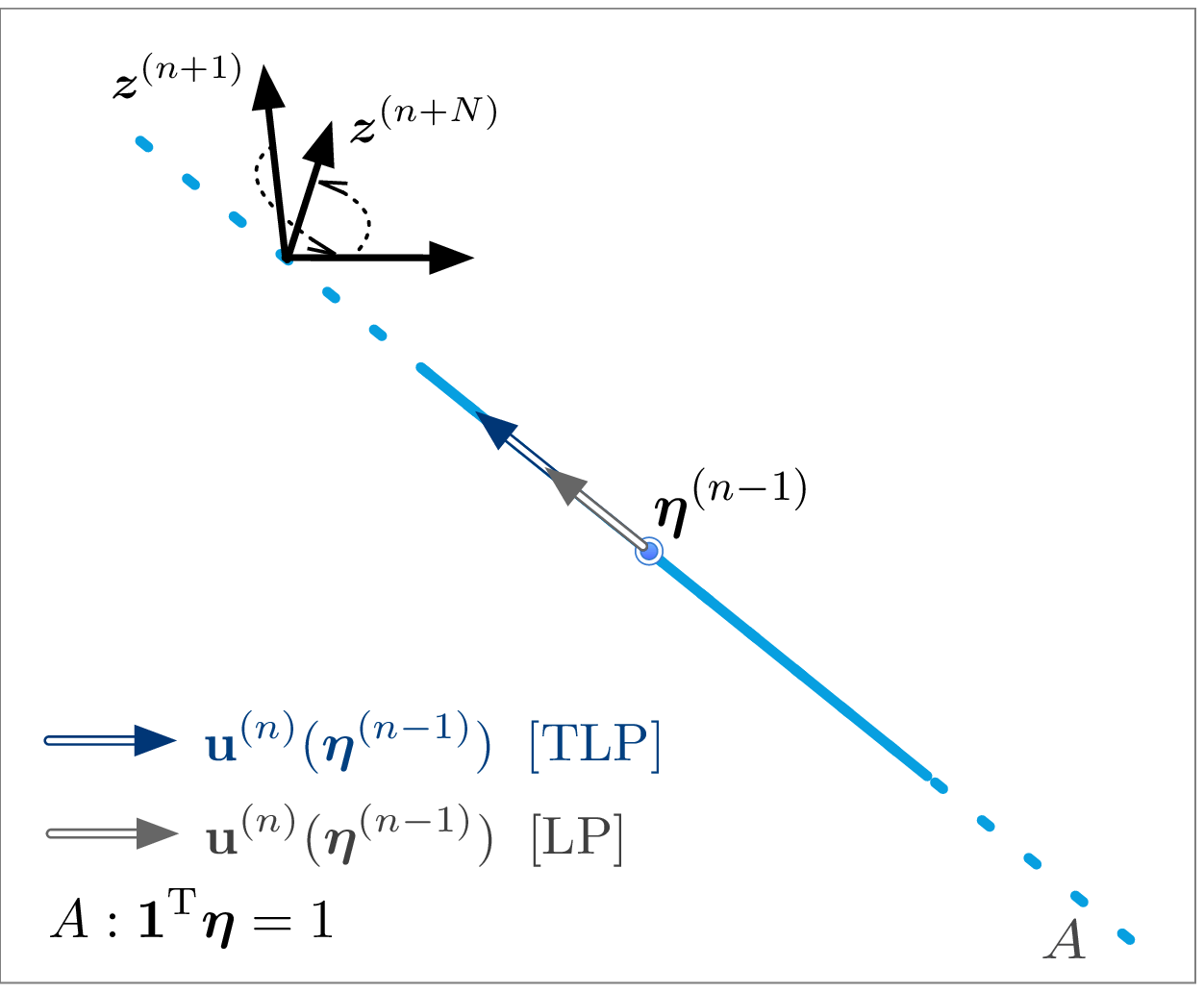} 
		\label{f:PopularityOscillate}}
	\vspace{2mm}
	\caption{Illustration of the relation between the replacement schemes, the instantaneous STF $\mathbf{u}^{(n)}(\boldsymbol{\eta}^{(n-1)})$, and the state cache hit probability $\boldsymbol{z}^{(n+1)}$.}\label{f:AnaNumPcol} \vspace{-2mm}
\end{figure*}

\subsection{Observations}

The following observations can be made from the preceding analysis of the relation between the instantaneous STF and the difference in cache hit probability.~\footnote{Accurate prediction of content popularity is assumed for the case of LP and TLP.}

\begin{itemize}
	\item From eq.~\eqref{e:ul_RR}, eq.~\eqref{e:difHitRate_n+1_2}, and eq.~\eqref{e:alpha}, it can be seen that the parameter $\phi$ is only a scaling factor in $d^{(n+1)}_\gamma$ in the case of RR. Specifically, whether $d^{(n+1)}_\gamma$ is negative or not is jointly decided by $\boldsymbol{\upsilon}^{(n+1)}$, $\boldsymbol{\upsilon}^{(n)}$, and $\boldsymbol{\eta}^{(n-1)}$. The parameter $\phi$ can scale $d^{(n+1)}_\gamma$ but does not have any impact on its sign. This explains the result in \cite{J_JGaoPartI2018} that $\phi$ impacts on the convergence speed but not the steady state under constant content popularity. 
	\item Four cases of instantaneous STF $\mathbf{u}^{(n)}(\boldsymbol{\eta}^{(n-1)})$ and $\boldsymbol{z}^{(n+1)}$ are illustrated in Fig.~\ref{f:ulcase12}~and~Fig.~\ref{f:ulcase34}. In each single replacement, both LP and TLP drive the SCP $\boldsymbol{\eta}$ towards a direction that increases $(\boldsymbol{z}^{(n+1)})^\mathrm{T}\boldsymbol{\eta}$, i.e., $(\boldsymbol{z}^{(n+1)})^\mathrm{T}\boldsymbol{\eta}^{(n+1)} \geq (\boldsymbol{z}^{(n+1)})^\mathrm{T}\boldsymbol{\eta}^{(n)}$, where $\boldsymbol{z}^{(n+1)} = \mathbf{C}_\mathrm{s}^\mathrm{T} \boldsymbol{\upsilon}^{(n+1)}$. Therefore, only case~2 in Fig.~\ref{f:ulcase12} and case~3 in Fig.~\ref{f:ulcase34} are possible for LP and TLP while all four cases can occur for RR and LRU. Moreover, TLP drives $\boldsymbol{\eta}$ towards the direction that increases $(\boldsymbol{z}^{(n+1)})^\mathrm{T}\boldsymbol{\eta}$ the fastest, which is a resemblance to the steepest gradient in optimization. This explains the result in the first part that TLP converges faster than LP under constant content popularity. 
	\item Under time-varying content popularity, LP and TLP may not effectively trace the varying content popularity depending on the pattern of variation. Specifically, if $\boldsymbol{\upsilon}^{(n)}$ varies so that $\boldsymbol{z}^{(n)}$ changes along a straight path over time, as shown in Fig.~\ref{f:PopularityStraight}, then LP and TLP can still trace the content popularity well, and TLP should outperform LP. An example of such scenario is when popularity concentrates so that the most popular contents become even more popular over time. 
	\item If $\boldsymbol{\upsilon}^{(n)}$ varies so that $\boldsymbol{z}^{(n)}$ changes fast and randomly in an area, as shown in Fig.~\ref{f:PopularityOscillate}, then LP and TLP may not trace the content popularity well, and TLP can perform worse than LP. An example of such scenario is when content popularity varies drastically over time so that the most popular set of contents rapidly changes. 
\end{itemize}


\section{Numerical Examples}\label{s:Simu}

\subsection{Instantaneous STF under Time-varying Content Popularity}

\begin{figure}[t]
	\centering 
	\subfloat[RR.]
	{\includegraphics[angle=0,width=0.40\textwidth]{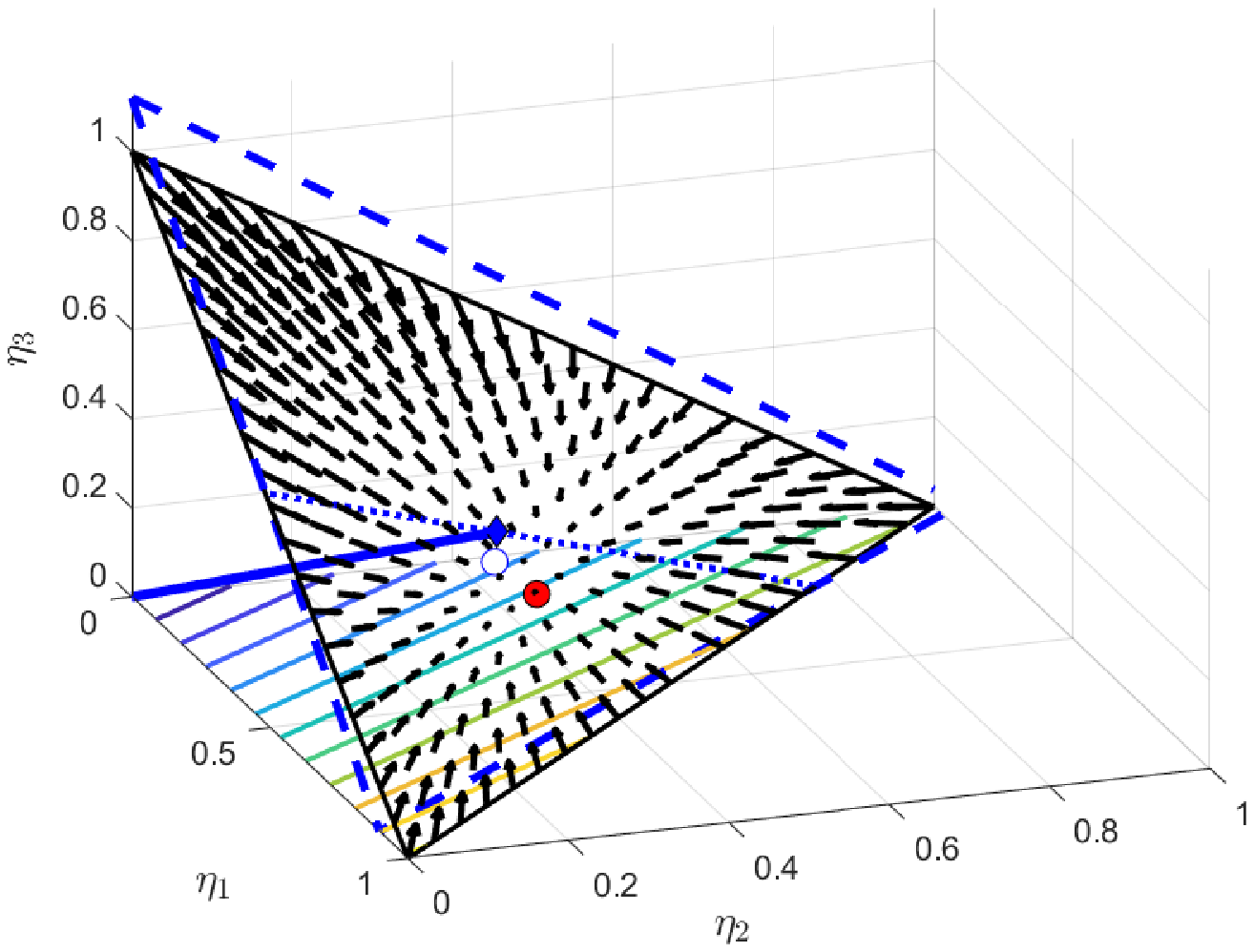}\label{f:RR_Varying}}
	\hspace{-1mm} 
	\subfloat[LP, example 1.]
	{\includegraphics[angle=0,width=0.40\textwidth]{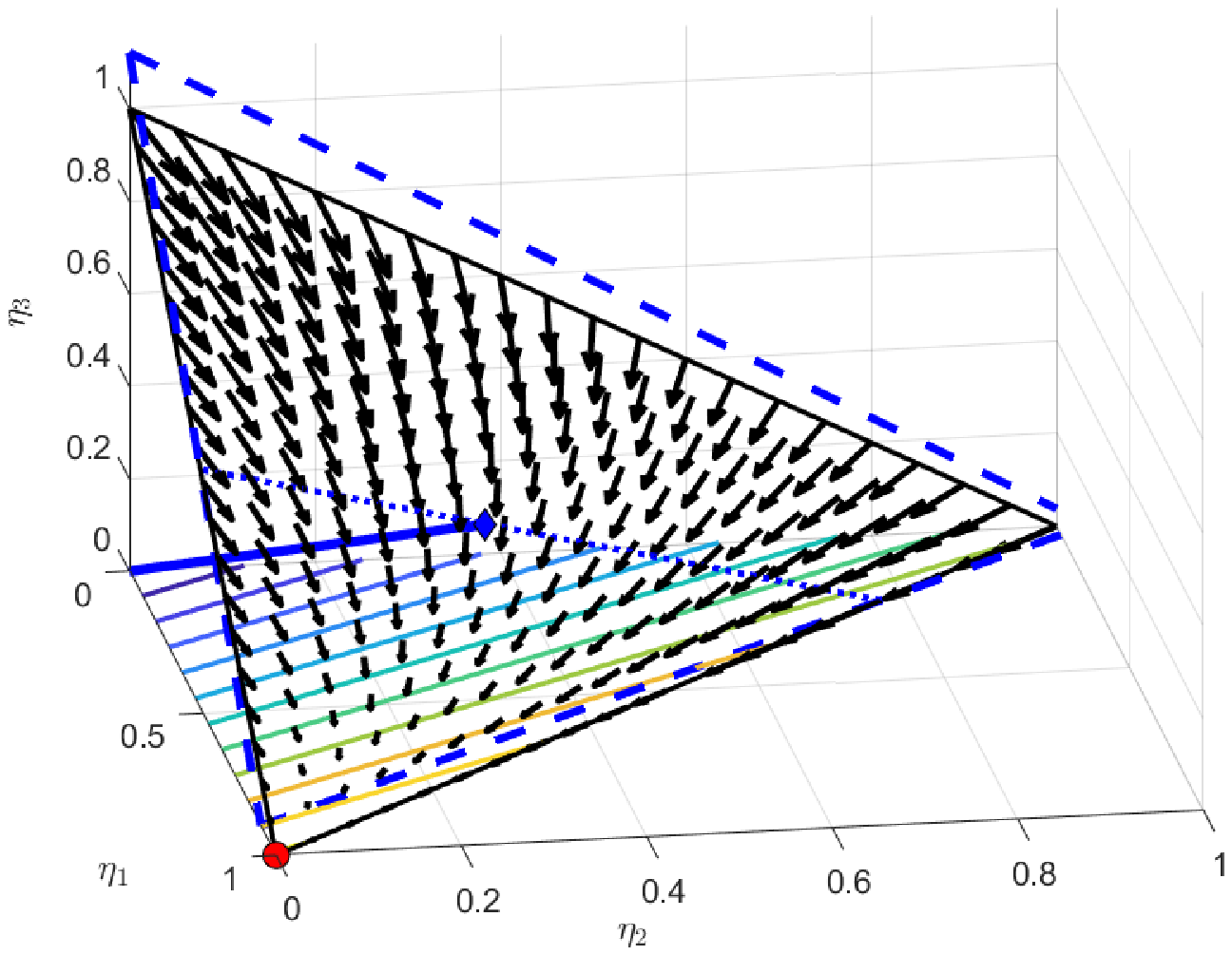}\label{f:LES_Varying1}}
	\hspace{1mm} 
	\subfloat[LP, exapmle 2.]
	{\includegraphics[angle=0,width=0.40\textwidth]{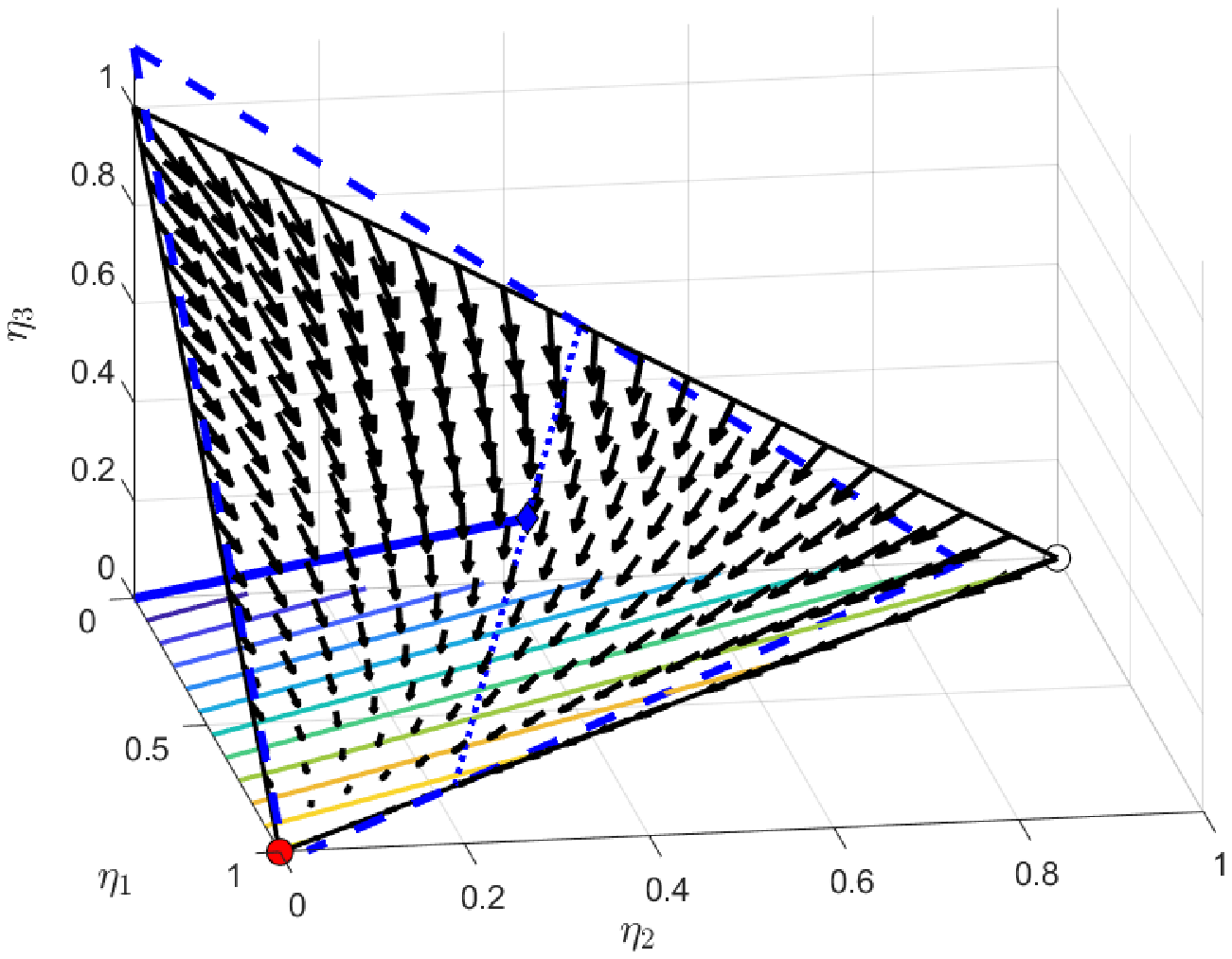}\label{f:LES_Varying2}}
	\vspace{2mm}
	\caption{Instantaneous STF and its impact on the instantaneous cache hit probability at the next request.}\label{f:field3DVarying} \vspace{-2mm}
\end{figure}

Fig.~\ref{f:field3DVarying} demonstrates the instantaneous STF under time-varying content popularity and further illustrates Fig.~\ref{f:OneStepImrpovement} using RR and LP as examples. Similar to the first part of this two-part paper, we use 3-D STFs for illustrations.   

Fig.~\ref{f:RR_Varying} shows the case under RR. The content popularity at the $n$th and $(n+1)$th requests are $\boldsymbol{\upsilon}^{(n)} = [0.46,  0.30,  0.24]^\mathrm{T}$ and $\boldsymbol{\upsilon}^{(n+1)} = [0.4, 0.35, 0.25]^\mathrm{T}$, respectively. The solid circle with red filling shows where the steady state would be if the content popularity were fixed and equal to $\boldsymbol{\upsilon}^{(n)}$. The hollow circle shows where the stationary state would be if the content popularity were fixed and equal to $\boldsymbol{\upsilon}^{(n+1)}$. The black triangular area with solid edges represents the state transition domain. The black arrows demonstrate the direction and strength of the STF at the instant of the $n$th request and the corresponding locations in the state transition domain. The colored straight lines in the x-y plane show the contour of the cache hit probability in the state transition domain. The solid straight line from the origin $(0,0,0)$ to the diamond marker in the STF are specified by the vector $\mathbf{C}_\mathrm{s} \boldsymbol{\upsilon}^{(n+1)}$. Denote the SCP vector $\boldsymbol{\eta}$ at the diamond marker as $\bar{\boldsymbol{\eta}}^{(n)}$. The dashed triangle in blue represents the intersection of the plane $(\boldsymbol{\upsilon}^{(n+1)})^\mathrm{T} \mathbf{C}_\mathrm{s} (\boldsymbol{\eta} - \bar{\boldsymbol{\eta}}^{(n)}) = 0$ with the 3 planes $\eta_1= 0$, $\eta_2 = 0$, and $\eta_3 = 0$. The dotted line represents the intersection of the plane $(\boldsymbol{\upsilon}^{(n+1)})^\mathrm{T} \mathbf{C}_\mathrm{s}  (\boldsymbol{\eta} - \bar{\boldsymbol{\eta}}^{(n)}) = 0$ with the state transition domain. 

From Fig.~\ref{f:RR_Varying}, the effect of the $n$th replacement, given the replacement scheme of RR and the above change of content popularity from $\boldsymbol{\upsilon}^{(n)}$ to  $\boldsymbol{\upsilon}^{(n+1)}$, can be observed. Specifically, given any SCP, i.e, a point in the state transition domain, if the arrow representing the instantaneous STF at that point can be scaled such that it crosses the dotted line from below to above, the $n$th replacement yields a smaller cache hit probability at the $(n+1)$th request compared with no replacement. By contrast, if the arrow can be scaled such that it crosses the dotted line from above to below, the $n$th replacement yields a larger cache hit probability at the $(n+1)$th request. If the arrow is in parallel with the dotted line, the $n$th replacement has no impact on the cache hit probability at the $(n+1)$th request.

Fig.~\ref{f:LES_Varying1} shows the first of two examples with LP. The content popularity $\boldsymbol{\upsilon}^{(n)}$ and $\boldsymbol{\upsilon}^{(n+1)}$ are the same as in Fig.~\ref{f:RR_Varying}. In this example, the change in the content popularity is not significant so that the state which caches the most popular contents does not change. As a result, the stationary state if the content popularity is fixed and equal to $\boldsymbol{\upsilon}^{(n)}$ and that if the content popularity is fixed and equal to $\boldsymbol{\upsilon}^{(n+1)}$ are identical and shown by a solid circle in the figure. The dashed triangle, solid straight line, and dotted line illustrate the same objects or variables as in Fig.~\ref{f:RR_Varying}, respectively. The effect of the $n$th replacement on the cache hit probability at the $(n+1)$th request at any SCP point in the state transition domain can be observed from Fig.~\ref{f:LES_Varying1} following the same method described in the preceding paragraph. In this example, the arrow at any point (except the stationary point) can be scaled such that it crosses the dotted line with the arrow head below the line. As a result, a replacement after request $n$ based on LP always increases the cache hit probability at the $(n+1)$th request (except at $\boldsymbol{\eta} = [1, 0 ,0]$). This example corresponds to the scenario of varying content popularity which drives $\boldsymbol{z}^{(n)}$ along a somewhat straight path, as shown in Fig.~\ref{f:PopularityStraight}.

Fig.~\ref{f:LES_Varying2} shows the second example with LP. The content popularity $\boldsymbol{\upsilon}^{(n)}$ is the same as in Fig.~\ref{f:RR_Varying} and Fig.~\ref{f:LES_Varying1}, while $\boldsymbol{\upsilon}^{(n+1)} = [0.4, 0.25, 0.35]^\mathrm{T}$. The solid and the hollow circles show the stationary states in the cases when the content popularity is fixed and equal to $\boldsymbol{\upsilon}^{(n)}$ and $\boldsymbol{\upsilon}^{(n+1)}$, respectively. At any SCP point, if the arrow can be scaled such that it crosses the dotted line from right to left, the $n$th replacement yields a smaller cache hit probability at the $(n+1)$th request compared with no replacement. By contrast, if the arrow can be scaled such that it crosses the dotted line from left to right, the $n$th replacement yields a larger cache hit probability at the $(n+1)$th request. In this example, a replacement after request $n$ based on LP may either increase or decrease the cache hit probability at the $(n+1)$th request. This example corresponds to the scenario of varying content popularity which leads to a randomly changing $\boldsymbol{z}^{(n)}$, as shown in Fig.~\ref{f:PopularityOscillate}.

\subsection{Cache Hit Ratio under Time-varying Content Popularity}

\begin{figure}[t]
	\centering 
	\subfloat[Cache hit ratio versus $t_0^{\max}$.]
	{\includegraphics[angle=0,width=0.38\textwidth]{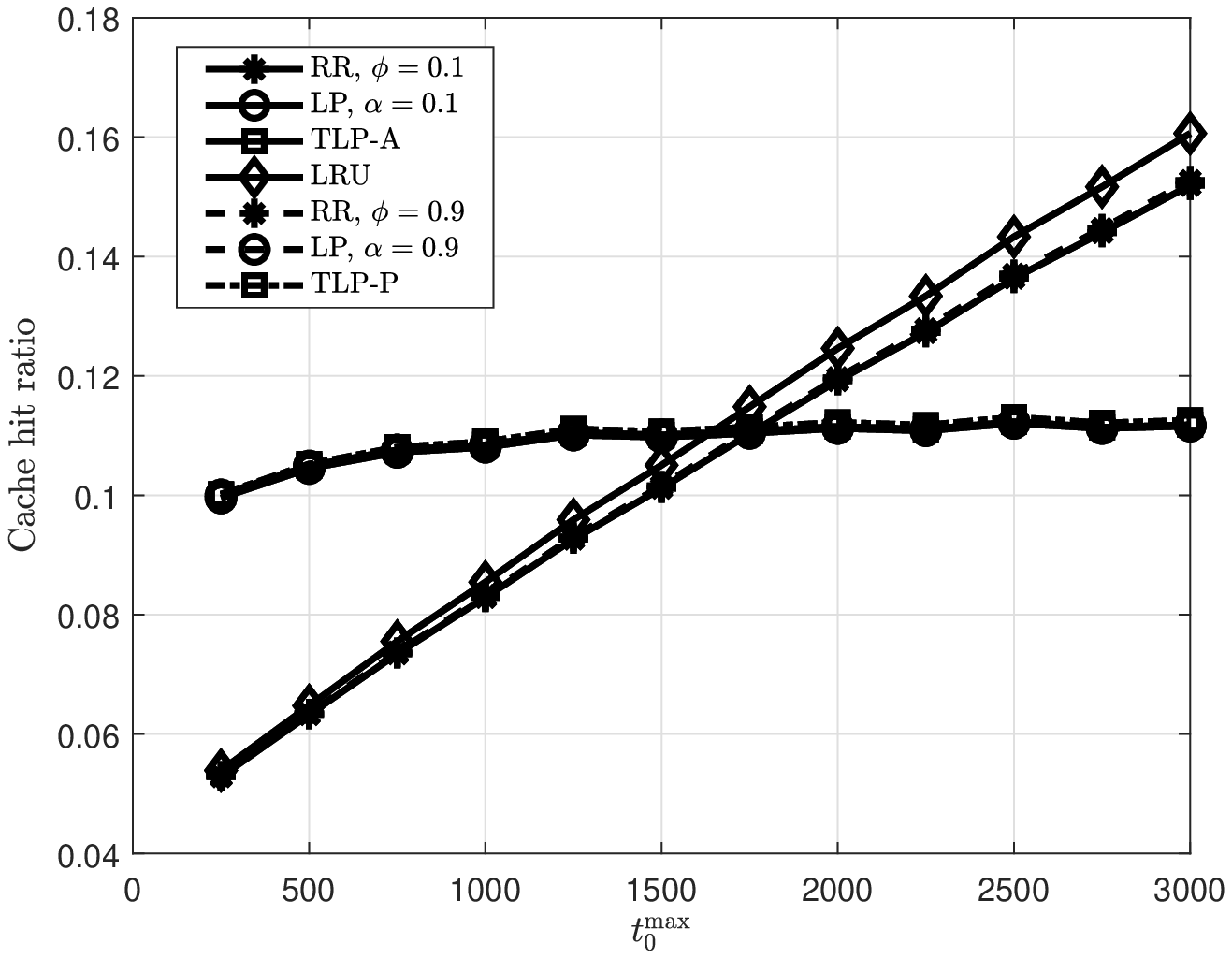}\label{f:HitRatioVary1}}
	\hspace{-1mm} 
	\subfloat[Request instants for 40 out of 1000 contents when $t_0^{\max} = 250$.]
	{\includegraphics[angle=0,width=0.38\textwidth]{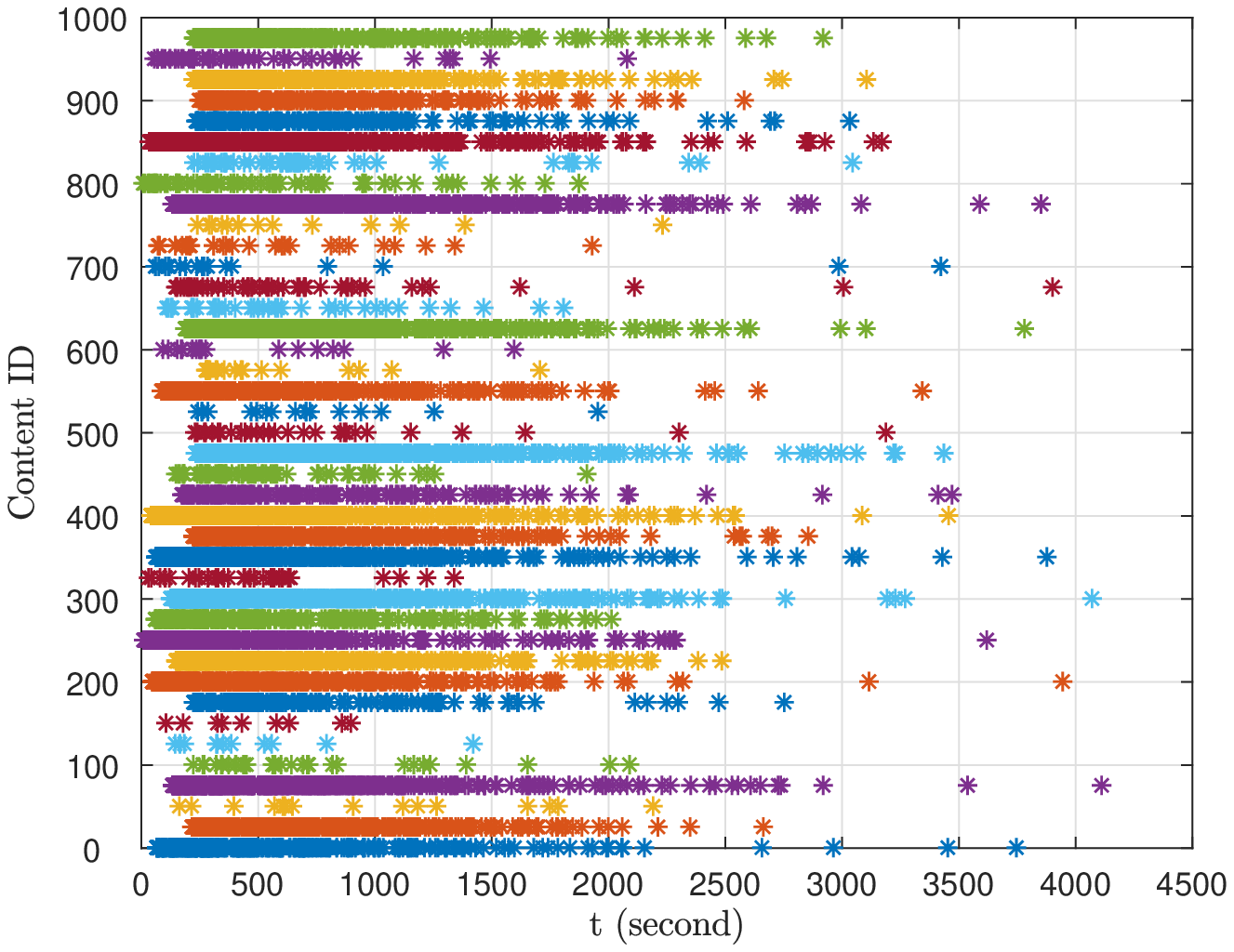}\label{f:HitRatioVary1Request1}}
	\hspace{-1mm} 
	\subfloat[Request instants for 40 out of 1000 contents in one round when $t_0^{\max} = 2500$.]
	{\includegraphics[angle=0,width=0.38\textwidth]{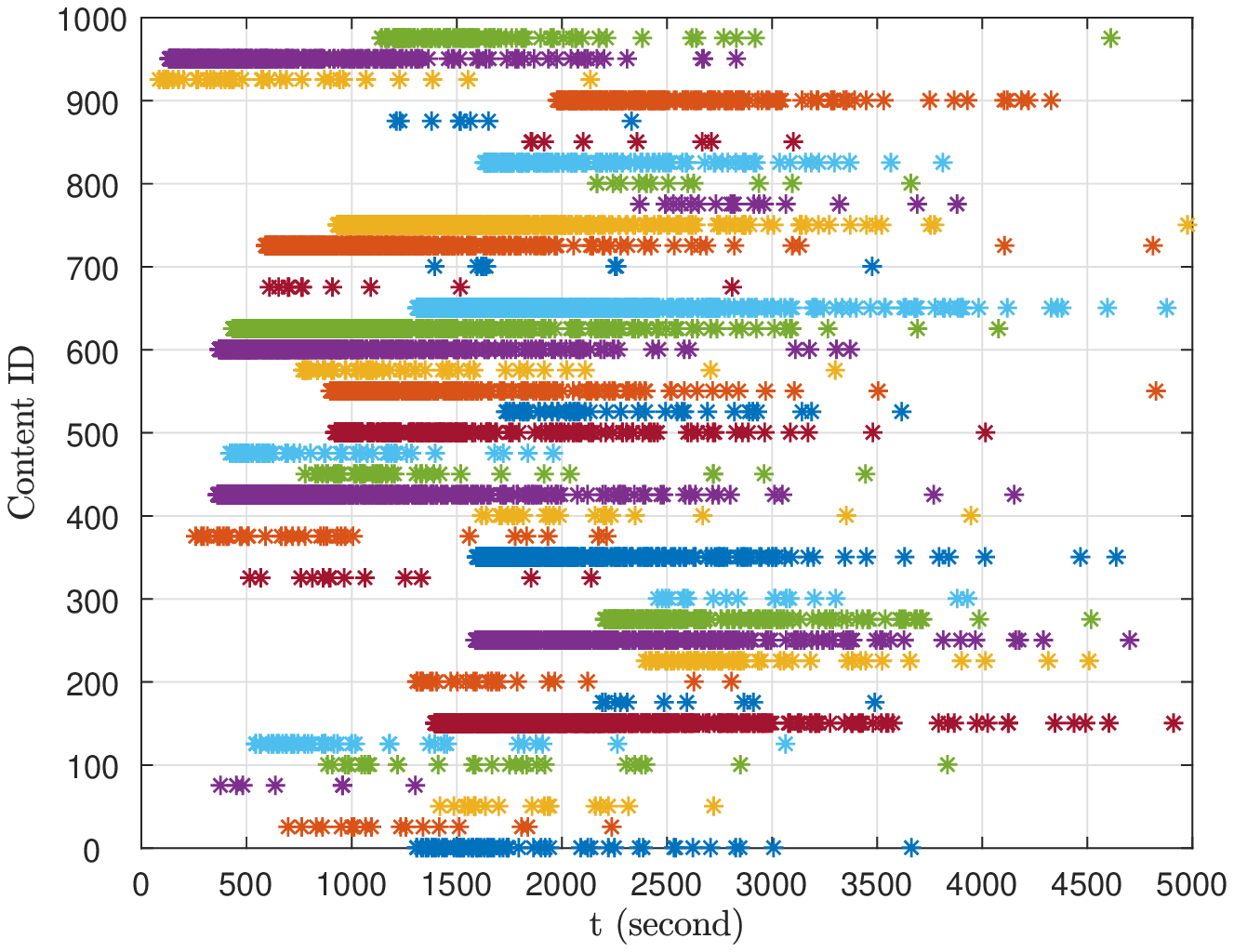}\label{f:HitRatioVary1Request2}}
	\vspace{0mm}
	\caption{Cache hit ratio under shot noise model.}\label{f:HitRatioShotNoise} \vspace{-2mm}
\end{figure}

In the second set of examples, the cache hit ratio of the considered cache replacement schemes under time-varying content popularity is demonstrated.

First, the cache hit ratio is demonstrated when the time-varying content popularity is generated using the shot noise model \cite{J_STraverso2013}. Specifically, the request for content $l$ follows a time-inhomogeneous
Poisson process with the instantaneous rate at time $t$ given by:
\begin{align}\label{e:ShotNoiseInstRate}
y_l(t) = \left\{
\begin{array}{ll}
A_l b_l \exp^{-b_l(t-t_{l, 0})}  , & \text{if}\;\, t \geq t_{l,0}\\
0, \quad &\text{otherwise}
\end{array}
\right.
\end{align}
Accordingly, requests for content $l$ start occurring from $t_{l, 0}$. The parameter $A_l$ limits the maximum request rate of content $l$. For content $l$, an allocation of $A_l$ over time is given by an exponential distribution with rate parameter $b_l$.  It follows that contents have different life-span and entrance time. The entrance time $t_{l, 0}$ is uniformly generated in $[0, t_0^{\max}]$, and $A_l$ is uniformly generated in $[A_l^{\min}, A_l^{\max}]$. For RR, we test two cases, $\phi = 0.9$ and $\phi = 0.1$. A larger $\phi$ results in more frequent content replacements and higher sensitivity to the changes in the content popularity. Similarly, for LP, we test two cases, i.e., $\alpha = 0.9$ and $\alpha = 0.1$.

In the first example with shot noise model, the number of contents $N_\mathrm{c}$ is set to 1000 and the cache size $L$ is set to $15$. A duration with 5000 seconds from $t= 0$ to $t = 5000$ is considered. The parameters $A_l^{\min}$ and $A_l^{\max}$ are set to 10 and 1000, respectively. Fig.~\ref{f:HitRatioVary1} shows the resulting cache hit ratio of the considered replacement schemes versus $t_0^{\max}$. Each data point in Fig.~\ref{f:HitRatioVary1} is averaged over 200 rounds of simulations for the considered 5000 seconds duration. For LP and TLP, accurate prediction of content popularity is assumed. It can be seen from the Fig.~\ref{f:HitRatioVary1} that LP and TLP have a significant advantage over RR and LRU when $t_0^{\max}$ is small (i.e., $t_0^{\max} \leq 1000$). However, RR and LRU are much better than LP and TLP when $t_0^{\max}$ becomes large. 

The content request time instants for 40 out of the 1000 contents~\footnote{Specifically, the contents whose content ID is a multiple of 25 are selected.} in the case when $t_0^{\max} = 250$ and $t_0^{\max} = 2500$ are plotted in Figs.~\ref{f:HitRatioVary1Request1}~and~\ref{f:HitRatioVary1Request2}, respectively. Colors are used to distinguish the requests for different contents. Each asterisk in Figs.~\ref{f:HitRatioVary1Request1}~and~\ref{f:HitRatioVary1Request2} represents a request, with its x~and~y coordinates specifying the corresponding request time instant and the content ID, respectively. It can be seen from Figs.~\ref{f:HitRatioVary1Request1}~and~\ref{f:HitRatioVary1Request2} that, when $t_0^{\max}$ becomes large, the set of available contents can vary significantly over time. This has two effects on the cache hit ratio. On one hand, the cache hit ratio should increase as the number of simultaneous available contents can be smaller when $t_0^{\max}$ is large. On the other hand,  due to the property of the instantaneous rate given by eq.~\eqref{e:ShotNoiseInstRate}, the maximum instantaneous request rate of any content occurs when the content just becomes available. If follows that the varying set of available contents when $t_0^{\max}$ is large can lead to frequent and abrupt change of content popularity over time, as illustrated in Fig.~\ref{f:PopularityOscillate} and Fig.~\ref{f:LES_Varying2}. Since LP and TLP exploit the content popularity information in a greedy manner (i.e., maximizing the cache hit ratio based on the current content popularity information), the second effect can hinder the cache hit ratio, and the combined impact of the above two effects yields an almost steady cache hit ratio of LP and TLP in Fig.~\ref{f:HitRatioVary1}. By contrast, the cache hit ratio of RR and LRU increases with $t_0^{\max}$ as the result of the first effect while the second effect has no significant impact as RR and LRU do not rely on the instantaneous content popularity information.

\begin{figure}[t]
	\centering 
	\subfloat[Cache hit ratio versus $t_0^{\max}$.]
	{\includegraphics[angle=0,width=0.38\textwidth]{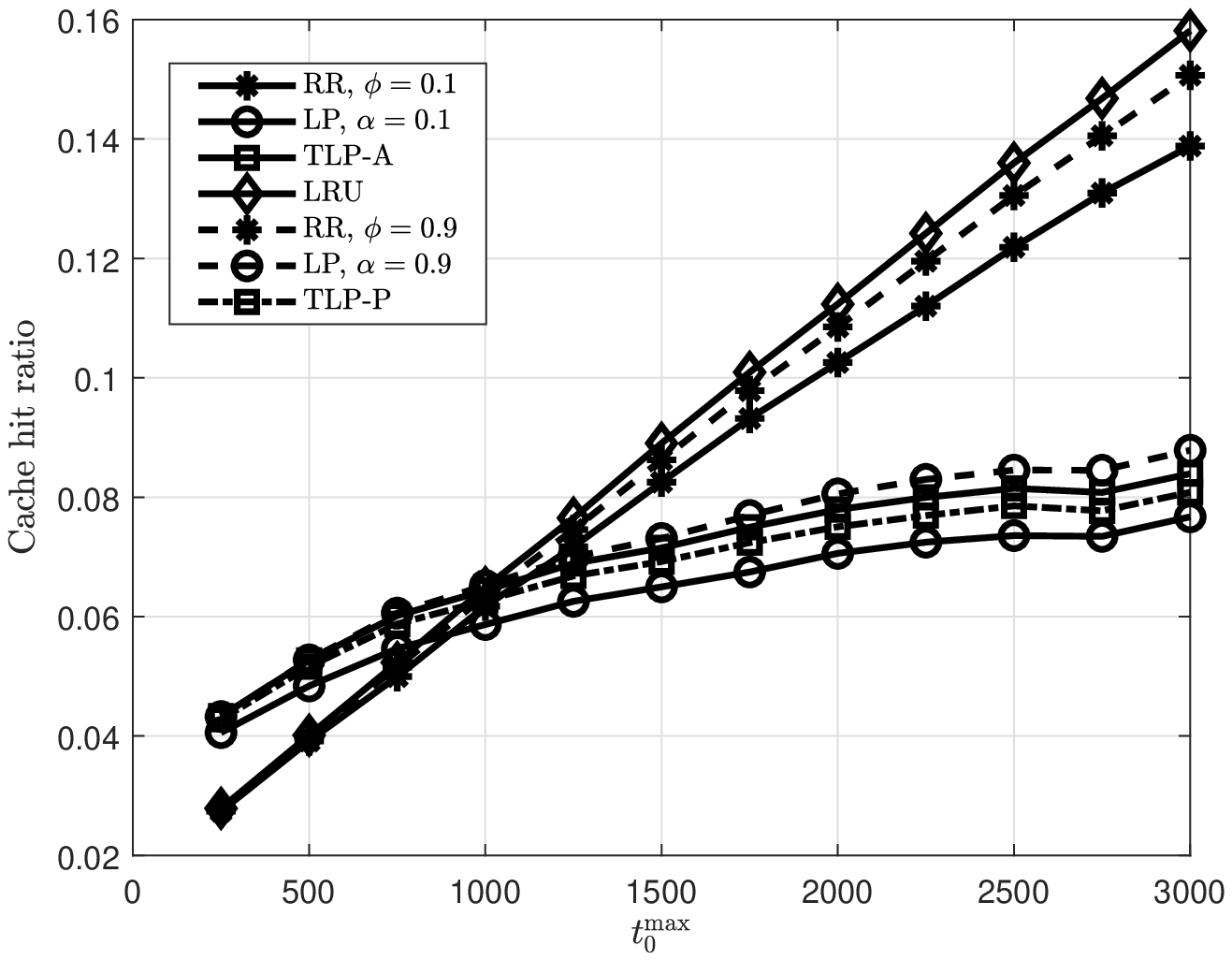}\label{f:HitRatioVary2}}
	\hspace{-1mm} 
	\subfloat[Request instants for 40 out of 2000 contents in one round when $t_0^{\max} = 2500$.]
	{\includegraphics[angle=0,width=0.38\textwidth]{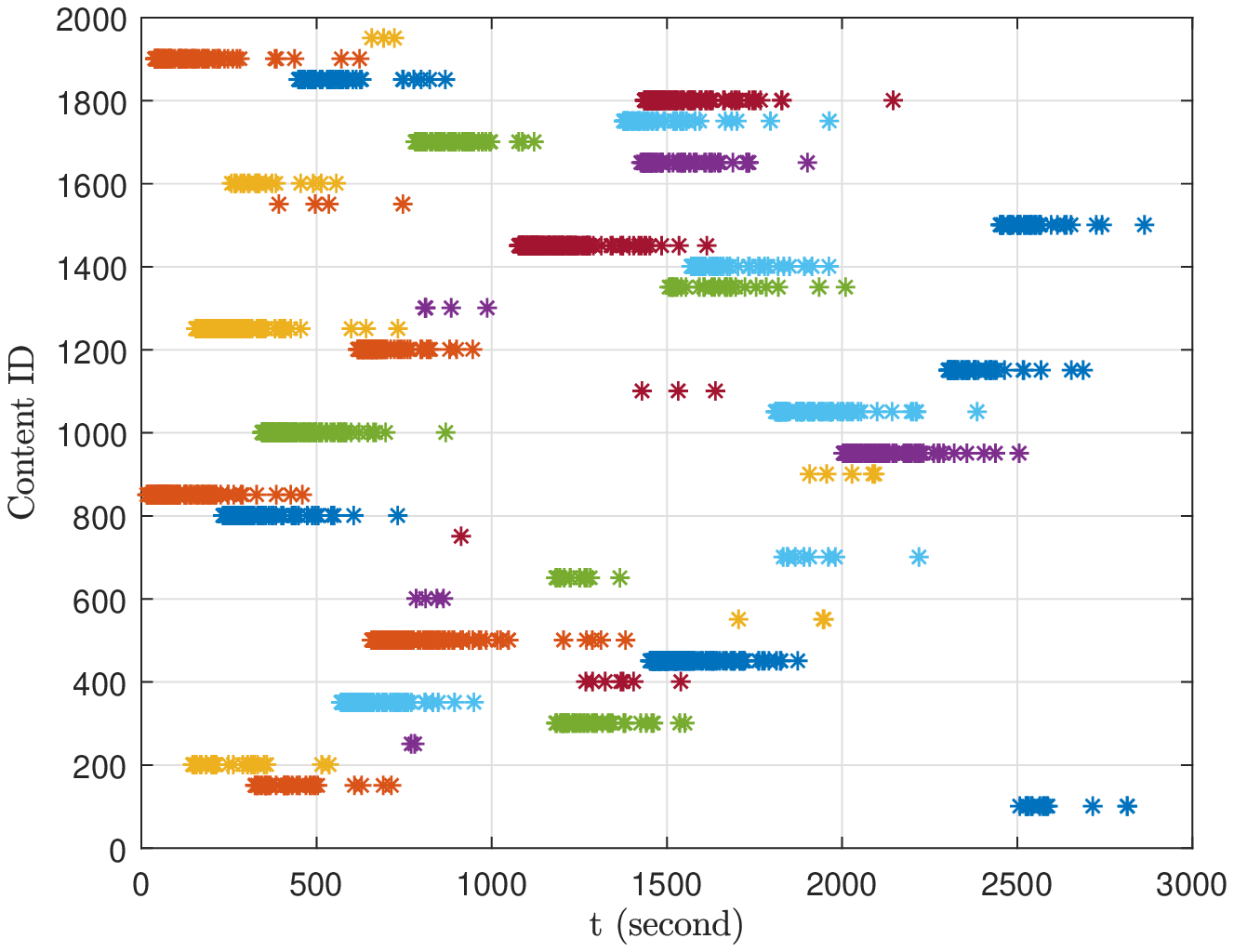}\label{f:HitRatioVary2Request}}
	\vspace{0mm}
	\caption{Cache hit ratio under shot noise model, short life-span.}\label{f:HitRatioShotNoise2} \vspace{-2mm}
\end{figure}

In the second example with shot noise model, $N_\mathrm{c}$ is increased from 1000 to 2000, and $A_l^{\max}$ and $A_l^{\min}$ are decreased from 1000 to 200 and from 10 to 1, respectively. The average content life-span also becomes shorter. Fig.~\ref{f:HitRatioVary2} shows the resulting cache hit ratio versus $t_0^{\max}$, while the request time instants for 40 out of the 2000 contents when $t_0^{\max} = 2500$ is plotted in Fig.~\ref{f:HitRatioVary2Request}. Comparing Fig.~\ref{f:HitRatioVary2} with Fig.~\ref{f:HitRatioVary1}, three observations can be made. First, the cache hit ratio in Fig.~\ref{f:HitRatioVary2} becomes lower for all schemes when $t_0^{\max}= 0$, as a result of $N_\mathrm{c}$ increasing to 2000. Second, the effect of $\phi$ and $\alpha$ on the performance of RR and LP, respectively, becomes obvious in Fig.~\ref{f:HitRatioVary2}. This is because a larger $\phi$ or $\alpha$ allows for a faster adaption to new content requests, which is important now that the number of requests for each content decreases significantly. Third, RR and LRU begin to outperform LP and TLP from a smaller $t_0^{\max}$ in Fig.~\ref{f:HitRatioVary2} compared to that in Fig.~\ref{f:HitRatioVary1}, and the performance gap between the two groups becomes larger. This is because the combination of active contents and their popularity varies even more rapidly compared with the case in Fig.~\ref{f:HitRatioVary1}, as a result of a larger $N_\mathrm{c}$ and shorter content life span. The result in Fig.~\ref{f:HitRatioVary2} shows that exploiting the instantaneous content popularity information in a content replacement scheme is not necessarily beneficial for increasing the cache hit ratio even if such information is predicted perfectly. This is because the usefulness of the instantaneous content popularity information depends on how rapidly the content popularity changes. This example corresponds to the case illustrated in Fig.~\ref{f:PopularityOscillate}.

\begin{figure}[t]
	\centering 
	\subfloat[Cache hit ratio versus $t_0^{\max}$.]
	{\includegraphics[angle=0,width=0.38\textwidth]{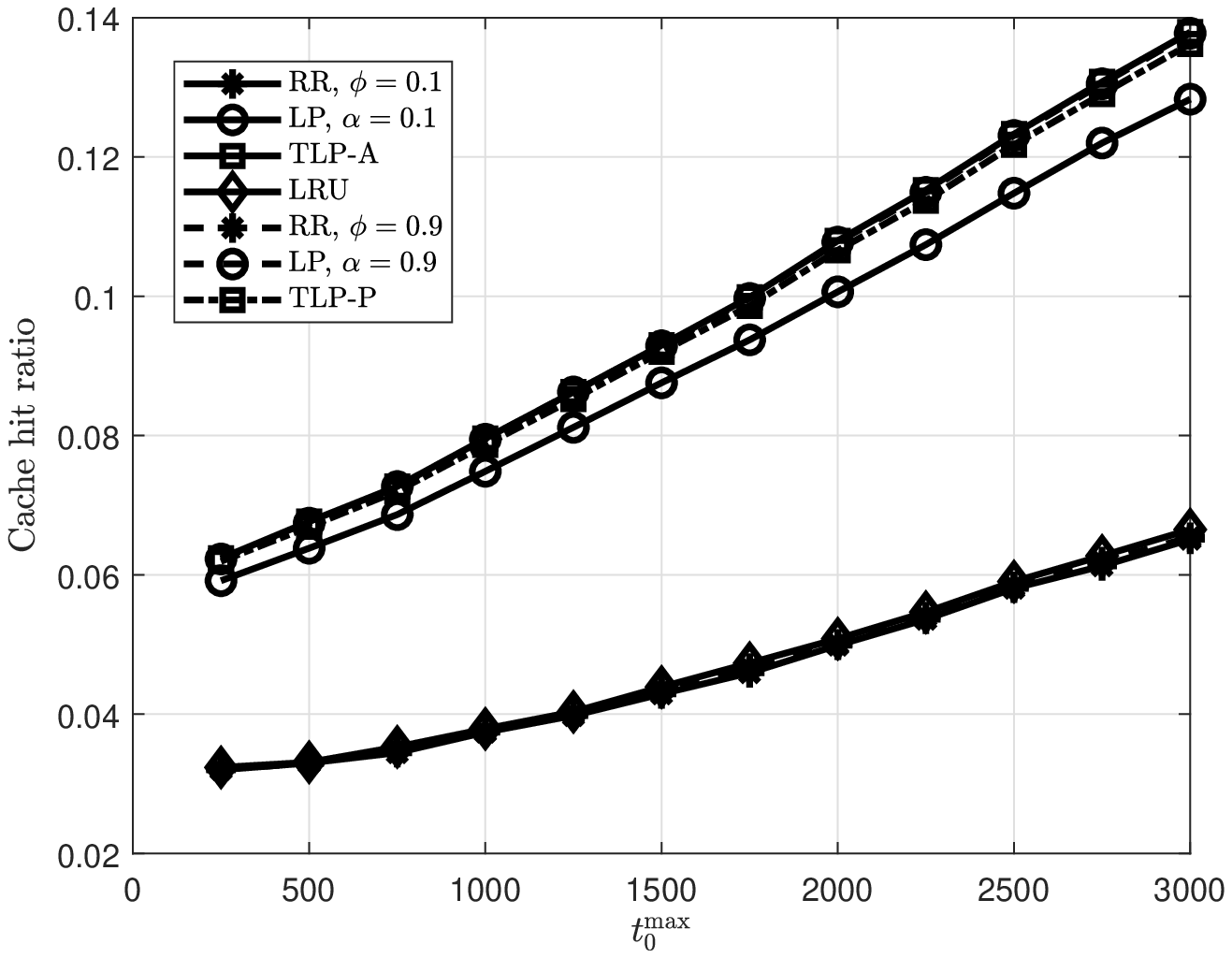}\label{f:HitRatioVary3}}
	\hspace{-1mm} 
	\subfloat[Request instants for 40 out of 1000 contents  in one round  when $t_0^{\max} = 2500$.]
	{\includegraphics[angle=0,width=0.38\textwidth]{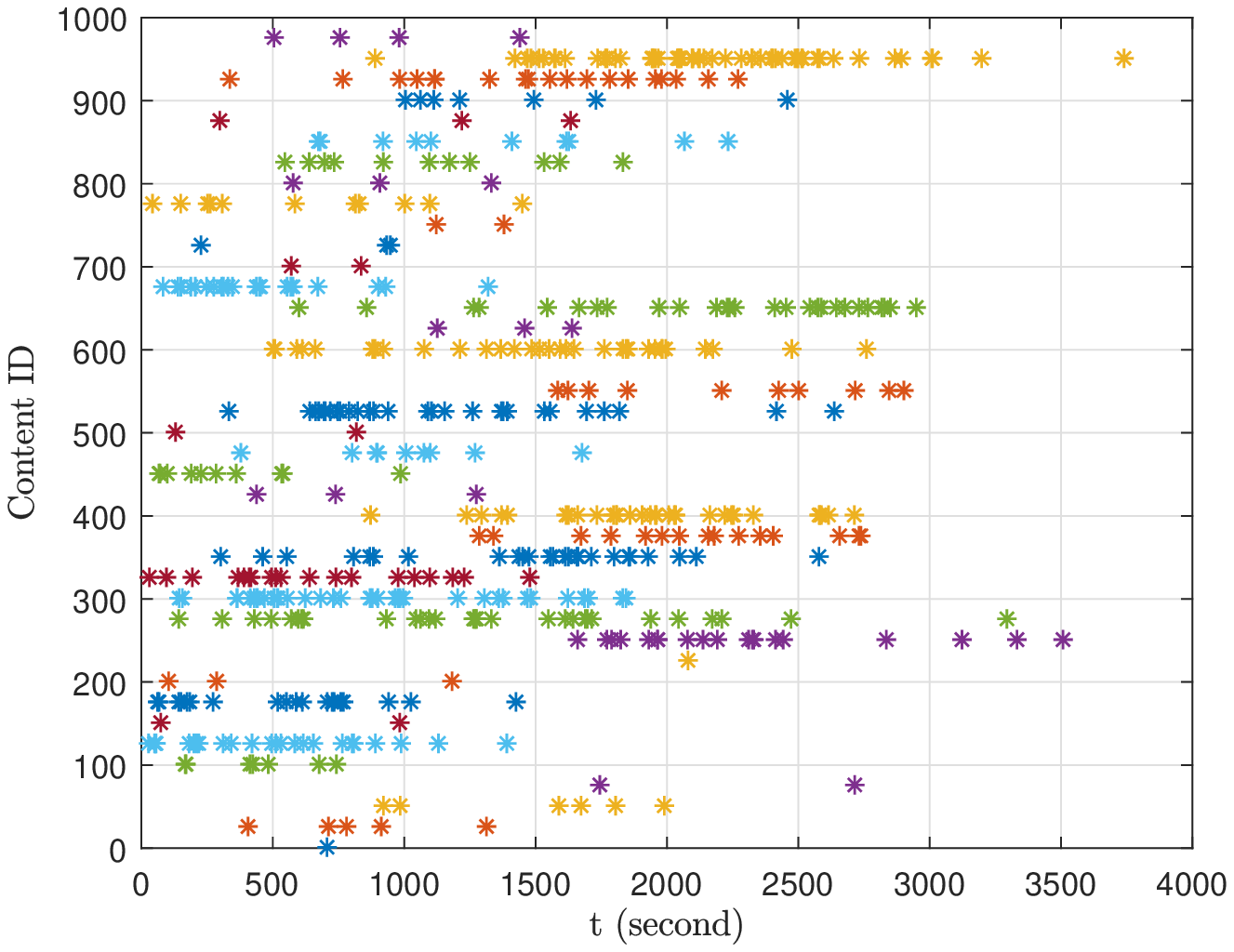}\label{f:HitRatioVary3Request}}
	\vspace{0mm}
	\caption{Cache hit ratio under time-inhomogeneous Poisson process represented by eq.~\eqref{e:GaussianInstRate}.}\label{f:HitRatioShotNoise3} \vspace{-2mm}
\end{figure}

Fig.~\ref{f:HitRatioShotNoise3} shows the cache hit ratio with a time-varying content popularity model different from eq.~\eqref{e:ShotNoiseInstRate}. Specifically, the request for content $l$ follows a time-inhomogeneous Poisson process with the instantaneous rate at time $t$ given by:
\begin{align}\label{e:GaussianInstRate}
y_l(t) =A_l \frac{1}{\sqrt{2\pi} \sigma}\exp^{- \frac{(t - t_{l, 0})}{2 \sigma^2}}.
\end{align}
The parameter $t_{l, 0}$ is no longer the entrance time of instant $l$ in eq.~\eqref{e:GaussianInstRate}. However, $t_{l, 0}$ in both eq.~\eqref{e:ShotNoiseInstRate} and eq.~\eqref{e:GaussianInstRate} corresponds to the time instant of the peak instantaneous request rate for content $l$. Similarly, $t_{l, 0}$ is uniformly generated in $[0, t_0^{\max}]$, and $A_l$ is uniformly generated in $[A_l^{\min}, A_l^{\max}]$. 

In this simulation, $N_\mathrm{c}$ is set to 1000 and the cache size $L$ is set to $15$. A duration with 5000 seconds from $t= 0$ to $t = 5000$ is considered. The parameters $A_l^{\min}$ and $A_l^{\max}$ are set to 1 and 50, respectively. Fig.~\ref{f:HitRatioVary3} shows the resulting cache hit ratio of the considered replacement schemes versus $t_0^{\max}$, in which each data point is averaged over 200 rounds of simulations. Accurate prediction of content popularity is again assumed for LP and TLP. The request time instants for 40 out of the 1000 contents when $t_0^{\max} = 2500$ is plotted in Fig.~\ref{f:HitRatioVary3Request}. It can be seen that Fig.~\ref{f:HitRatioVary3} shows a very different result when compared with Fig.~\ref{f:HitRatioVary1} or Fig.~\ref{f:HitRatioVary2}. Specifically, LP and TLP always perform better that RR and LRU in Fig.~\ref{f:HitRatioVary3}, and the performance gap between the two groups increases with $t_0^{\max}$. This is because that, unlike the abrupt and frequent variations introduced by eq.~\eqref{e:ShotNoiseInstRate}, the instantaneous rate model in eq.\eqref{e:GaussianInstRate} leads to smooth and graduate variations in the content popularity. As a result, the instantaneous content popularity at any instant can be close to the instantaneous content popularity for a number of subsequent requests. Therefore, 
the greedy maximization of the cache hit ratio based on the current content popularity information used by LP and TLP can benefit the cache hit ratio for both the immediate next request and also subsequent requests. Consequently, the LP and TLP outperform RR and LRU due to the exploration of the instantaneous content popularity information in such case. This example corresponds to the case illustrated in Fig.~\ref{f:PopularityStraight}.

\section{Conclusion}

We have extended the study of dynamic caching via STF to the case of time-varying content popularity. In our analysis, we have focused on developing the model and methodology without assuming a specific pattern of change in content popularity. The results have demonstrated the impact of varying popularity on the STF and the performance of replacement schemes in the general case. Further extensions can be conducted by incorporating a specific model of time-varying content popularity. In our simulations, we have adopted different models of varying popularity, and the numerical results have been shown to be consistent with the observations from the analysis. 

Through the two parts of this paper, we have provided a novel perspective and developed methods for studying cache replacement in the vector space of SCP using STF. It has been demonstrated that the design of replacement schemes is essentially the design of STF and that the knowledge of content popularity is beneficial only if exploited properly, depending on the pattern of change in the content popularity. As there are many open issues, especially in the case of time-varying content popularity, the results of this paper have been developed in the effort of inspiring the analysis or design of cache replacement schemes for various specific problems and scenarios.

\vspace{2mm}

\appendix

\subsection{Proof of Lemma~\ref{l:JointPLRU}}\label{p:JointPLRU}
Suppose that the LRU content at the $n$th request is content $q^\star$, and the most recent request for $q^\star$ is the $(n - w)$th request. It must hold that $w \geq L$, and all requests from $(n-w+1)$th request to the $(n-1)$th request must be for contents $l \in \mathcal{C}_k\backslash\{q^\star\}$. Denote the $N_\mathrm{c}$ contents in $l \in \mathcal{C}_k\backslash\{q^\star\}$ as $k(1, \bar{q}^\star), \dots k(L- 1, \bar{q}^\star)$. 
To allocate the total number of $w-1$ requests (i.e., from the $(n-w+1)$th request to the $(n-1)$th request) to the $L- 1$ contents in $l \in \mathcal{C}_k\backslash\{q^\star\}$, there are $P_w = \binom{w-1}{L- 1}$ different different allocations, without considering the order of requests, that guarantees at least one request for each content. Denote the number of requests for content $k(i, \bar{q}^\star)$ in the $j$th combination as $T(k, i, j, \bar{q}^\star)$, where $i \in \{1, \dots, L-1\}$ and $j \in \{1, \dots, P_w\}$. It follows that: 
\begin{align}
\sum_{1 =1}^{L-1} T(k, i, j, \bar{q}^\star)= w-1, \forall j.
\end{align}
Then, considering the order of request, the number of different ordered allocations are:
\begin{align}
U_w = \sum\limits_{j = 1}^{P_w} \prod\limits_{i = 1}^{L- 1} \binom{w- \! 1 \! - a_{k, i, j, \bar{q}^\star}}{T(k, i, j, \bar{q}^\star)}.
\end{align}
in which
\begin{align}
a_{k, i, j, \bar{q}^\star}
=   \left\{
\begin{array}{ll}
0,  &\text{if} \;\;  i = 1  \\
\sum\limits_{y = 1}^{i-1} T(k, y, j, \bar{q}^\star), &\text{if}\;\; i \geq 2.
\end{array}
\right.
\end{align}

Denote the set of request instants for content $k(i, \bar{q}^\star)$ in the $u$th ordered combination as $\mathcal{T}(k, i, u, \bar{q}^\star)$, where $i \in \{1, \dots, L-1\}$ and $u \in \{1, \dots, U_w\}$. It follows that:
\begin{align}
\bigcup \limits_{i =1}^{L-1} \mathcal{T}(k, i, u, \bar{q}^\star) = \{n - 1, \dots, n-w+1\}, \forall u.
\end{align}
Accordingly, the joint probability that the current state is $k$, content $q^\star = e(k,m) \in \mathcal{C}_k$ is the LRU content at the $n$th request, and the most recent request for the LRU is the $(n-w)$th request is given by 
eq.~\eqref{e:lemmajointP}. \hfill$\blacksquare$

\subsection{Proof of Lemma~\ref{l:gammaSequence}}\label{p:gammaSequence}

The average cache hit probability from the 1st till the $n$th request is given by:
\begin{align}\label{e:gammaAvgDef}
\gamma_\textrm{avg} &= \frac{1}{n} \sum\limits_{t = 1}^{n} \left(\boldsymbol{\upsilon}^{(t)}\right)^\mathrm{T} \boldsymbol{\lambda}^{(t-1)} \nonumber \\
&= \frac{1}{n} \sum\limits_{t = 1}^{n} \left(\boldsymbol{\upsilon}^{(t)}\right)^\mathrm{T} \mathbf{C}_\mathrm{s} \boldsymbol{\eta}^{(t-1)}.
\end{align}
Using eq.~\eqref{e:etaSequenceDecomp} (and setting $n = 1$ and $N =  t- 1$ in eq.~\eqref{e:etaSequenceDecomp}), it holds that:
\begin{align}\label{e:etaSequenceDecomp2}
\boldsymbol{\eta}^{(t -1)} = \sum\limits_{t^\prime = 0}^{t-2} \boldsymbol{u}^{(t^\prime +1)}(\boldsymbol{\eta}^{(t^\prime)}) + \boldsymbol{\eta}^{(0)} 
\end{align} 
for any $t \geq 2$. Substituting eq.~\eqref{e:etaSequenceDecomp} into eq.~\eqref{e:gammaAvgDef}, it holds that:
\begin{align}
\gamma_\textrm{avg} =& \frac{1}{n} \sum\limits_{t = 2}^{n} \left(\boldsymbol{\upsilon}^{(t)}\right)^\mathrm{T} \mathbf{C}_\mathrm{s} \bigg(\sum\limits_{t^\prime = 0}^{t-2} \boldsymbol{u}^{(t^\prime +1)}(\boldsymbol{\eta}^{(t^\prime)}) + \boldsymbol{\eta}^{(0)} \bigg) \nonumber \\
& + \frac{1}{n} \left(\boldsymbol{\upsilon}^{(1)}\right)^\mathrm{T} \mathbf{C}_\mathrm{s} \boldsymbol{\eta}^{(0)} \nonumber \\
= & \frac{1}{n} \sum\limits_{t = 2}^{n} \left(\boldsymbol{\upsilon}^{(t)}\right)^\mathrm{T} \mathbf{C}_\mathrm{s} \bigg(\sum\limits_{t^\prime = 0}^{t-2} \boldsymbol{u}^{(t^\prime +1)}(\boldsymbol{\eta}^{(t^\prime)}) \bigg) \nonumber \\
& + \frac{1}{n} \sum\limits_{t = 1}^{n} \left(\boldsymbol{\upsilon}^{(t)}\right)^\mathrm{T} \mathbf{C}_\mathrm{s} \boldsymbol{\eta}^{(0)}  \nonumber\\
= & \frac{1}{n} \sum\limits_{t = 2}^{n} \left(\boldsymbol{\upsilon}^{(t)}\right)^\mathrm{T} \mathbf{C}_\mathrm{s} \bigg(\sum\limits_{t^\prime = 0}^{t-2} \boldsymbol{u}^{(t^\prime +1)}(\boldsymbol{\eta}^{(t^\prime)}) \bigg) \nonumber \\
& + \left(\frac{1}{n}\sum\limits_{t = 1}^{n} \boldsymbol{\upsilon}^{(t)}\right)^\mathrm{T}  \mathbf{C}_\mathrm{s} \boldsymbol{\eta}^{(0)},
\end{align}
which leads to eq.~\eqref{e:gammaSequenceLemma}.  \hfill $\blacksquare$

\subsection{Proof of Theorem~\ref{t:dGammaTheo}}\label{p:dGammaTheo}

As $\bar{\upsilon}_l$ is defined such that $\boldsymbol{\eta}^{(n-1)}$ would be the steady state if the content request probabilities were time-invariant and equal to $\{\bar{\upsilon}_l\}$. It follows that:
\begin{align}\label{e:virtualLambdaEqualities}
\sum\limits_{l \in \mathcal{C}} \bar{\upsilon}_l \boldsymbol{\Theta}_l \boldsymbol{\eta}^{(n-1)} = \boldsymbol{\eta}^{(n-1)}.
\end{align} 
Based on eq.~\eqref{e:u_decomp} and eq.~\eqref{e:virtualLambdaEqualities}, it holds that:
\begin{align}
\boldsymbol{\eta}^{(n)} - \boldsymbol{\eta}^{(n-1)} & = \sum\limits_{l \in \mathcal{C}} \left(\upsilon_l^{(n)} - \bar{\upsilon}_l\right) \boldsymbol{\Theta}_l \boldsymbol{\eta}^{(n-1)} \nonumber \\
& = \sum\limits_{l \in \mathcal{C}} \left(\upsilon_l^{(n)} - \bar{\upsilon}_l\right)  (\boldsymbol{\eta}^{(n-1)} + \mathbf{u}_l^{(n)} ) \nonumber \\
& = \sum\limits_{l \in \mathcal{C}} \left(\upsilon_l^{(n)} - \bar{\upsilon}_l\right) \mathbf{u}_l^{(n)}, 
\end{align}
where the last equality uses the fact that $\sum\limits_{l \in \mathcal{C}} \left(\upsilon_l^{(n)} - \bar{\upsilon}_l\right) = 0$. 

Substituting the above equation into eq.~\eqref{e:difHitRate_n+1} gives
\begin{align}
d^{(n+1)}_\gamma = \left(\boldsymbol{\upsilon}^{(n+1)} \right)^\mathrm{T} \mathbf{C}_\mathrm{s}\sum\limits_{l \in \mathcal{C}} \left(\upsilon_l^{(n)} - \bar{\upsilon}_l\right) \mathbf{u}_l^{(n)}.
\end{align}   
Rearranging the above equation using eq.~\eqref{e:alpha} leads to eq.~\eqref{e:difHitRate_n+1_2}. \hfill $\blacksquare$

\subsection{Proof of Theorem~\ref{t:dBound}}\label{p:dBound}

The proof is based on the equality $d^{(n+1)}_\gamma = \left(\boldsymbol{\upsilon}^{(n+1)} \right)^\mathrm{T} \mathbf{C}_\mathrm{s}\boldsymbol{u}^{(n)}$ in eq.~\eqref{e:difHitRate_n+1}. The elements of the $N_\mathrm{c} \times 1$ vector $\mathbf{C}_\mathrm{s}\boldsymbol{u}^{(n)}$ are the changes in the caching probabilities of the $N_\mathrm{c}$ contents after the $n$th request and replacement. It is straightforward to see that the upper and lower bounds of $d^{(n+1)}_\gamma$ are decided by the maximum and minimum elements of $\mathbf{C}_\mathrm{s}\boldsymbol{u}^{(n)}$, respectively.

Given that contents are sorted based on their popularity at the $n$th request, the maximum element of $\mathbf{C}_\mathrm{s}\boldsymbol{u}^{(n)}$ for all cases but TLP-P corresponds to the case when content 1 is requested while it is being cached with probability zero. Using eq.~\eqref{e:u_decomp} and eqs.~\eqref{e:ul_RR} - \eqref{e:ul_LRU},  it can be seen that the maximum element of $\mathbf{C}_\mathrm{s}\boldsymbol{u}^{(n)}$ is $L \phi  \max\limits_l \{ \upsilon_l^{(n)} \}$, $\alpha  \max\limits_l \{ \upsilon_l^{(n)} \}$, $\max\limits_l \{ \upsilon_l^{(n)} \}$, and $\max\limits_l \{ \upsilon_l^{(n)} \}$ for RR, LP, TLP-A, and LRU, respectively. For the case of TLP-P, it holds that
\begin{align}
\hat{d}^{(n+1)}_\gamma  \leq \max\limits_l \{ \upsilon_l^{(n)} \} \cdot \max\limits_l \{\tilde{\upsilon}_l^{(n+1)} - \tilde{\upsilon}_{N_\mathrm{c}}^{(n+1)}\}. 
\end{align}

For all cases but TLP-P, the minimum of $\mathbf{C}_\mathrm{s}\boldsymbol{u}^{(n)}$ corresponds to the following scenario: i). the state with the $L$ least popular contents is being cached with probability 1; and ii). a content not in the cache is requested. The change in the SCP of this state in the described scenario gives the minimum of $\mathbf{C}_\mathrm{s}\boldsymbol{u}^{(n)}$.

For RR, the change in the above SCP is given by
\begin{align}
\check{d}^{(n+1)}_\gamma = - \phi \sum\limits_{l = 1}^{N_\mathrm{c}-L} \upsilon_l^{(n)} \geq - \phi, 
\end{align}
where the inequality is based on the approximation that the summation of request probabilities of all but the $L$ least popular contents should be close to 1.  

For LP,  the change is given by  
\begin{align}
\check{d}^{(n+1)}_\gamma &= - \alpha \sum\limits_{l = 1}^{N_\mathrm{c}-L} \upsilon_l^{(n)} \frac{ \tilde{\upsilon}_l^{(n+1)} - \tilde{\upsilon}_{N_\mathrm{c}}^{(n+1)} }{ \sum\limits_{q = N_\mathrm{c}-L+1 }^{N_\mathrm{c}}  (\tilde{\upsilon}_l^{(n+1)} - \tilde{\upsilon}_{q}^{(n+1)} )}  \nonumber \\
& \geq - \alpha \sum\limits_{l = 1}^{N_\mathrm{c}-L} \upsilon_l^{(n)} \geq - \alpha.
\end{align}

For both TLP-A and LRU, the state will change as long as the requested content is not in the cache. Therefore, the aforementioned change is given by  
\begin{align}
\check{d}^{(n+1)}_\gamma &= -\sum\limits_{l = 1}^{N_\mathrm{c}-L} \upsilon_l^{(n)} \geq -1.
\end{align}

For TLP- P, assuming that the $L$ least popular contents at the $n$th request remain to be the least popular at the $(n+1)$th request, the change is given by  
\begin{align}
\check{d}^{(n+1)}_\gamma &= -\sum\limits_{l = 1}^{N_\mathrm{c}-L} \upsilon_l^{(n)} \left(\tilde{\upsilon}_l^{(n+1)} - \tilde{\upsilon}_{N_\mathrm{c}}^{(n+1)}\right) \nonumber \\
& \geq  - \sum\limits_{l = 1}^{N_\mathrm{c}-L} \upsilon_l^{(n)} \tilde{\upsilon}_l^{(n+1)} \geq - \sum\limits_{l = 1}^{N_\mathrm{c}}  \upsilon_l^{(n)}\tilde{\upsilon}_l^{(n+1)}.
\end{align}
This completes the proof of Theorem~\ref{t:dBound}. \hfill $\blacksquare$




{\footnotesize
\begin{thebibliography}{99}

\bibitem{M_XWang2014}
X. Wang, M. Chen, T. Taleb, A. Ksentini, and V. C. M. Leung, ``Cache in the air: exploiting content caching and delivery techniques for 5G systems,'' \textit{IEEE Commun. Mag.}, vol.~52, no.~2, pp.~131--139, Feb.~2014.

\bibitem{J_JSAC_SZhang2018}
S. Zhang, W. Quan, J. Li, W. Shi, P. Yang, and X. Shen, ``Air-Ground Integrated Vehicular Network Slicing With Content Pushing and Caching,'' \textit{IEEE J. Sel. Areas Commun.}, vol.~36, no.~9, pp.~2114--2127, Sept.~2018.

\bibitem{C_JGao2018}
J. Gao, L. Zhao, and L. Sun, ``Probabilistic Caching as Mixed Strategies in Spatially-Coupled Edge Caching,'' in \textit{Proc. 29th Biennial Symp.  Commun.}, Toronto, Canada, 2018.

\bibitem{J_SMuller2017}
S. M\"{u}ller, O. Atan, M. van der Schaar, and A. Klein, ``Context-Aware Proactive Content Caching With Service Differentiation in Wireless Networks,'' \textit{IEEE Trans. Wireless Commun.}, vol.~16, no.~2, pp.~1024--1036, Feb.~2017.

\bibitem{J_PYang2019}
P. Yang, N. Zhang, S. Zhang, L. Yu, J. Zhang, and X. Shen, ``Content Popularity Prediction Towards Location-Aware Mobile Edge Caching,''  \textit{IEEE Trans. Multimedia}, vol.~21, no.~4, pp.~915--929, Apr.~2019.

\bibitem{J_JGaoTMC2018}
J.~Gao, S.~Zhang, L.~Zhao, and X.~Shen, ``The Design of Dynamic Probabilistic Caching with Time-Varying Content Popularity,'' submitted to {\it IEEE Trans. Mobile Comput.}, under review.

\bibitem{J_KLi2018}
K. Li, C. Yang, Z. Chen, and M. Tao, ``Optimization and Analysis of Probabilistic Caching in $N$ -Tier Heterogeneous Networks,''  \textit{IEEE Trans. Wireless Commun.},  vol.~17, no.~2, pp.~1283--1297, Feb.~2018.

\bibitem{M_GPaschos2016}
G. Paschos, E. Bastug, I. Land, G. Caire, and M. Debbah, ``Wireless Caching: Technical Misconceptions and Business Barriers,''  \textit{IEEE Commun. Mag.}, vol.~54, no.~8, pp.~16--22, Aug.~2016.


\bibitem{J_GSPaschos2018}
G. S. Paschos, G. Iosifidis, M. Tao, D. Towsley, and G. Caire, ``The Role of Caching in Future Communication Systems and Networks,''  \textit{IEEE J. Sel. Areas Commun.}, vol.~36, no.~6, pp.~1111--1125, June~2018.

\bibitem{J_SAzimi2018}
S. M. Azimi, O. Simeone, A. Sengupta, and R. Tandon, ``Online Edge Caching and Wireless Delivery in Fog-Aided Networks With Dynamic Content Popularity,''  \textit{IEEE J. Sel. Areas Commun.}, vol.~36, no.~6, pp.~1189--1202, June 2018.


\bibitem{R_SJin1999}
S.~Jin and A.~Bestavros, ``Temporal Locality in Web Request Streams: Sources, Characteristics, and Caching Implications,'' Technical Report, BUCS-1999-014, Boston, USA, Oct. 1999.


\bibitem{C_MBusari2001}
M. Busari and C. Williamson, ``On the Sensitivity of Web Proxy Cache Performance to Workload Characteristics,'' in \textit{Proc. IEEE INFOCOM}, Anchorage, USA, 2001, pp. 1225--1234.

\bibitem{J_YZhou2015}
Y. Zhou, L. Chen, C. Yang, and D. M. Chiu, ``Video Popularity Dynamics and Its Implication for Replication,'' \textit{IEEE Trans. Multimedia}, vol.~17, no.~8, pp.~1273--1285, Aug.~2015.

\bibitem{C_VPhalke1995}
V. Phalke and B. Gopinath, ``An Inter-Reference Gap Model for Temporal Locality in Program Behavior,'' in \textit{Proc. ACM SIGMETRICS/PERFORMANCE Conf.},  Ottawa, Canada, May 1995, pp.~291--300.


\bibitem{J_STraverso2013}
S.~Traverso, M.~Ahmed, M.~Garetto, P.~Giaccone, E.~Leonardi, and S.~Niccolini, ``Temporal Locality in Today's Content Caching: Why It Matters and How to Model It,'' \textit{ACM SIGCOMM Comput. Commun. Rev.} vol.~43, no.~5, pp.~5--12. Nov.~2013.

\bibitem{J_STraverso2015}
S. Traverso, M. Ahmed, M. Garetto, P. Giaccone, E. Leonardi, and S. Niccolini, ``Unravelling the Impact of Temporal and Geographical Locality in Content Caching Systems,''  \textit{IEEE Trans. Multimedia}, vol.~17, no.~10, pp.~1839-1854, Oct.~2015.

\bibitem{J_MGaretto2016}
M.~Garetto, E.~Leonardi, and V.~Martina, ``A Unified Approach to the Performance Analysis of Caching Systems,''  \textit{ACM Trans. Model. Perform. Eval. Comput. Syst.}, vol.~1, no.~3, May 2016.

\bibitem{C_MGaretto2015}
M. Garetto, E. Leonardi, and S. Traverso, ``Efficient Analysis of Caching Strategies under Dynamic Content Popularity,'' in \textit{Proc. IEEE INFOCOM}, Kowloon, China, Apr./May~2015, pp.~2263--2271.

\bibitem{J_AJaleel2010}
A.~Jaleel, K.~B.~Theobald, S.~C.~Steely Jr., and J.~Emer, ``High Performance Cache Replacement Using Re-reference Interval Prediction (RRIP),'' \textit{ACM SIGARCH Comput. Archit. News}, vol.~38, no.~3, pp.~60--71, June 2010.

\bibitem{C_SLi2016}
S. Li, J. Xu, M. van der Schaar, and W. Li, ``Popularity-driven content caching,'' in \textit{Proc. IEEE INFOCOM}, San Francisco, USA, 2016.

\bibitem{C_NZhang2016}
N. Zhang, K. Zheng, and M. Tao, ``Using Grouped Linear Prediction and Accelerated Reinforcement Learning for Online Content Caching,'' in \textit{Proc. IEEE ICC Workshops}, Kansas City, USA, May~2018.

\bibitem{J_ASadeghi2018}
A. Sadeghi, F. Sheikholeslami, and G. B. Giannakis, ``Optimal and Scalable Caching for 5G Using Reinforcement Learning of Space-Time Popularities,'' \textit{IEEE J. Sel. Areas Signal Process.}, vol.~12, no.~1, pp.~180--190, Feb.~2018.

\bibitem{J_DApplegate2016}
D. Applegate, A. Archer, V. Gopalakrishnan, S. Lee, and K. K. Ramakrishnan, ``Optimal Content Placement for a Large-Scale VoD System,'' \textit{IEEE/ACM Trans Netw.}, vol.~24, no.~4, pp.~2114--2127, Aug.~2016.

\bibitem{J_BBharath2018}
B. N. Bharath, K. G. Nagananda, D. G\"{u}nd\"{u}z, and H. V. Poor, ``Caching With Time-Varying Popularity Profiles: A Learning-Theoretic Perspective,''  \textit{IEEE Trans. Commun.}, vol.~66, no.~9, pp.~3837--3847, Sept.~2018.

\bibitem{J_JGaoPartI2018}
J.~Gao, L.~Zhao, and X.~Shen, ``The Study of Caching via State Transition Field - the Case of Time-Invariant Popularity'', \textit{IEEE Trans. Wireless Commun.}, accepted.

%
%
%
%
%
%
%
%
%
%
%
%
%
%
%
%
%
%
%
%
%
%
%
%
%
%
%
%
%
%
%
%
%
%
%
%
%
%
%
%
%
%
%
%
%
%
%
%
%
%
%



\end{thebibliography} }
\end{document}